\documentclass[a4paper,11pt]{article}
\usepackage{jcappub} 

\usepackage[T1]{fontenc}
\usepackage{braket, xcolor, soul, cleveref,fullpage,tabularx, diagbox,hyperref}
\let\vec\mathbf

\newcommand{\beq}{\begin{equation}}
\newcommand{\eeq}{\end{equation}}
\newcommand{\bea}{\begin{eqnarray}}
\newcommand{\eea}{\end{eqnarray}}
\newcommand{\nn}{\nonumber\\}
\newcommand{\dd}{\text{d}}
\newcommand{\dec}{\text{dec}}

\newcommand{\eff}{\text{eff}}

\newcommand{\Neff}{N_\text{eff}}
\newcommand{\Ndr}{N_\text{dr}}
\newcommand{\Geff}{G_\text{eff}}
\newcommand{\lgeff}{\log_{10}(G_\text{eff}\,\text{MeV}^2)}
\newcommand{\lleff}{\log_{10}(\lambda_\text{eff})}

\newcommand{\mk}{\mathbf{k}}
\newcommand{\mq}{\mathbf{q}}
\newcommand{\ml}{\mathbf{l}}
\newcommand{\avgrate}{\left< \Gamma \right>}
\newcommand{\effrate}{\Gamma_\textrm{eff}}

\newcommand{\rdr}{R_{\rm dr}}
\newcommand{\TT}{{\rm TT}}
\newcommand{\TE}{{\rm TE}}
\newcommand{\EE}{{\rm EE}}
\newcommand{\mpl}{M_{\rm Pl}}

\newcommand{\es}[2] {\begin{equation} \label{#1} \begin{split} #2 \end{split} \end{equation}}
\newcommand{\D}{\mathrm{d}}
\renewcommand{\arraystretch}{1.2}

\title{Cosmological Constraints on Secluded Dark Radiation}

\author[a,b]{Jae Hyeok Chang,}
\author[c]{Peizhi Du,}
\author[d]{Subhajit Ghosh,}
\author[e,f]{Soubhik Kumar}

\affiliation[a]{Theory Division,
            Fermi National Accelerator Laboratory, Batavia, IL 60510, USA}
\affiliation[b]{Department of Physics,
            University of Illinois Chicago, Chicago, IL 60607, USA}
\affiliation[c]{Laboratory of Spin Magnetic Resonance, School of Physical Sciences, Anhui Province Key Laboratory of Scientific Instrument Development and Application, University of Science and Technology of China, Hefei 230026, China}
\affiliation[d]{Texas Center for Cosmology and Astroparticle Physics, Weinberg Institute, Department of Physics, The University of Texas at Austin, Austin, TX 78712, USA}
\affiliation[e]{Institute of Cosmology, Department of Physics and Astronomy, Tufts University, Medford, MA 02155, USA}
\affiliation[f]{Center for Cosmology and Particle Physics, Department of Physics,
New York University, New York, NY 10003, USA}

\emailAdd{jhchang@fnal.gov}
\emailAdd{dupeizhi@ustc.edu.cn}
\emailAdd{sghosh@utexas.edu}
\emailAdd{soubhik.kumar@tufts.edu}

\abstract{Dark radiation (DR) is ubiquitous in physics beyond the Standard Model (SM), and its interactions with the SM and dark matter (DM) lead to a variety of interesting effects on cosmological observables. However, even in scenarios where DR is `secluded', i.e., only gravitationally interacting with SM and DM, it can leave discernible signatures. We present a comprehensive study of four different types of DR: free-streaming, self-interacting (coupled), decoupling, and recoupling DR, and vary initial conditions to include both adiabatic and isocurvature perturbations. In addition to these properties, we also vary neutrino energy density, DR energy density, and the SM neutrino masses to perform a general analysis and study degeneracies among neutrino and DR properties. 
We derive constraints using the cosmic microwave background, large-scale structure, and supernova datasets.
We find no significant preference for physics beyond the $\Lambda$CDM model, but data exhibit interesting interplays between different physical quantities.
When the neutrino energy density is allowed to vary, we find that the cosmological dataset prefers massless free-streaming DR over massive neutrinos, leading to a significant relaxation of the neutrino mass bound.
Although we do not find any evidence of DR isocurvature, the data show support for a strong blue tilt of the isocurvature power spectrum. Our analysis also highlights the degeneracy of various DR parameters with the Hubble constant $H_0$ resulting in a mild relaxation of the $H_0$ tension.
}

\begin{document}
\begin{flushright}
\texttt{FERMILAB-PUB-25-0710-T-V\\UT-WI-31-3025}
\end{flushright}
\setcounter{tocdepth}{2}
\maketitle
\flushbottom

\section{Introduction}

Current cosmological data can be well-described by the standard $\Lambda$CDM model across a wide range of scales~\cite{Planck:2018vyg}. 
However, the $\Lambda$CDM model is incomplete because it cannot explain the origin and nature of dark matter (DM) and dark energy. Furthermore, the $\Lambda$CDM model is facing certain tensions between different measurements (e.g., the tension between the early-time and late-time measurements of the Hubble expansion rate today, $H_0$). Many well-motivated models of physics beyond the Standard Model (BSM) that can address these theoretical and observational puzzles predict additional light degrees of freedom that behave like radiation during the decoupling of the cosmic microwave background (CMB).
This kind of radiation is typically called dark radiation (DR).
From a particle physics perspective, DR can arise in various forms, including axion(-like particles)~\cite{Marsh:2015xka}, dark photons~\cite{Fabbrichesi:2020wbt}, and light sterile neutrinos~\cite{Dasgupta:2021ies}. DR also arises in models addressing the origin of the Higgs mass, such as Twin Higgs~\cite{Chacko:2005pe} and N-naturalness models~\cite{Arkani-Hamed:2016rle}.
In this study, we focus on a certain type of DR that only couples to the Standard Model (SM) particles and DM via gravity. We call this \textit{secluded} dark radiation. It is extremely difficult to probe secluded DR by terrestrial and astrophysical observations due to their negligible interactions with SM.
Precision cosmological measurements, however, provide a unique tool to test physics that is only gravitationally coupled to the SM. This is because DR can carry a significant fraction of energy density in the early Universe and thus leave characteristic imprints on cosmological observables.

Dark radiation can have various kinds of self-interactions depending on the microphysics of the dark sector. We classify them into four broad types by comparing their self-interaction rate  $\Gamma(z)$ to the Hubble expansion rate $H(z)$ which goes as $z^{2}$ in radiation domination (RD), where $z$ is the cosmic redshift.
\begin{itemize}
    \item Free-streaming DR (FDR): $\Gamma(z)\ll H(z)$ throughout cosmic history and this can arise from models with weakly coupled DR or no self-interactions. 
    \item  Coupled DR (CDR): $\Gamma(z)\gg H(z)$ throughout cosmic history, meaning that DR is always strongly coupled and thus behaves like a coupled fluid (see, e.g.,~\cite{Buen-Abad:2015ova}).
    \item Decoupling DR: the self-interaction rate has stronger temperature dependence (i.e., $\Gamma(z)\propto z^{n}$ with $n>2$ in RD) than the Hubble rate, so that $\Gamma(z)> H(z)$ at early times, but $\Gamma(z)< H(z)$ at late times. We will see that the CMB data constrain scenarios where this transition takes place prior to or around recombination. 
    This type of interaction can arise if DR is coupled to itself via a heavy mediator (similar to various neutrino interaction scenarios)~\cite{Cyr-Racine:2013jua, Archidiacono:2013dua,Forastieri:2017oma,Kreisch:2019yzn,Barenboim:2019tux,Das:2020xke,RoyChoudhury:2020dmd,Brinckmann:2020bcn,Das:2025asx}. 
    \item Recoupling DR: the self-interaction rate has milder temperature dependence (i.e., $\Gamma(z)\propto z^n$ with $n<2$ in RD) than the Hubble rate, so that $\Gamma(z)< H(z)$ at early times, but $\Gamma(z)> H(z)$ at late times, opposite to the case of decoupling DR. This type of interaction can arise from scalar DR that has quartic couplings~\cite{Brinckmann:2022ajr}, or if DR is coupled to itself via a light mediator (similar to neutrino self-interactions with a light mediator)~\cite{Chacko:2003dt,Hannestad:2004qu,Bell:2005dr,Sawyer:2006ju,Friedland:2007vv, Forastieri:2015paa,Forastieri:2019cuf,Escudero:2019gvw,Escudero:2021rfi,Escudero:2020dfa,Sandner:2023ptm}.
\end{itemize} 
As we will see in later sections, different types of DR leave distinct features in the CMB spectra. Current CMB data  already have the sensitivity to probe the differences of these kinds of DR, and probe the total energy density of DR, as well as its interaction strength.

In addition to the dynamics of DR, cosmological observables also depend on the initial conditions of primordial density perturbations. The standard choice is the \textit{adiabatic} initial conditions for which the relative fluctuations of various components of the Universe do not vary from place to place, and all the fluctuations can be described by the fluctuations of a single variable, such as time or temperature.
This is a natural consequence of `single-clock' inflation, where all density perturbations were determined by the fluctuation of a single degree of freedom, such as the inflaton.
However, this property of adiabaticity is rather delicate, especially in the context of secluded DR.
A natural way in which secluded DR can arise is if the SM gets reheated by the inflaton $\phi$ while a second field $\chi$ decays into DR, with $\chi$ and $\phi$ only gravitationally coupled.
The primordial perturbations of $\phi$ and $\chi$ need not be correlated.
This implies the perturbations of SM and DR will not be `aligned', in general.
For example, regions that have an overdensity in SM need not have an overdensity in DR.
This corresponds to the presence of DR \textit{isocurvature} perturbations~\cite{Ghosh:2021axu}.\footnote{For general discussions of isocurvature, see Ref.~\cite{Bucher:1999re,Wands:2000dp,Gordon:2000hv,Lyth:2002my,Malik:2004tf,Wands:2007bd}.}
This is similar to the more well-known scenarios of cold dark matter (CDM) isocurvature from axion or curvaton models (for reviews, see e.g.,~\cite{Marsh:2015xka,Mazumdar:2010sa}). DR isocurvature perturbations evolve differently compared to adiabatic ones, which results in different shapes of angular power spectra in the CMB. Therefore, precision CMB measurements have the potential to probe new physics that generates DR isocurvature perturbations.

The main goal of this work is to provide an up-to-date and comprehensive analysis of various secluded DR scenarios, discussed above, where there is no direct interaction between DR and SM or DM. 
We organize representative DR cases by four types of self-interaction strengths (FDR, CDR, decoupling, and recoupling DR), two initial conditions (adiabatic and isocurvature), together with the freedom of varying neutrino and DR energy density as well as the SM neutrino masses. Subsets of these cases have been studied in the literature: the four types of interactions with adiabatic initial conditions were studied, for example, in~\cite{Baumann:2015rya,Choi:2018gho,Blinov:2020hmc,Brinckmann:2022ajr,Saravanan:2025cyi}; DR isocurvature with FDR and CDR was analyzed in~\cite{Kawasaki:2011rc, Adshead:2020htj,Ghosh:2021axu,Elor:2023xbz, Buckley:2024nen,Buckley:2025zgh}. However, all these scenarios studied before had limited freedom in the neutrino sector, meaning either neutrino mass or energy density was fixed. Therefore, it is worthwhile to provide a more complete analysis of secluded DR scenarios with updated datasets, as well as including the effects of varying neutrino energy density and masses. This allows us to study parameter degeneracies between the DR and neutrino sector in detail. Furthermore, from a microphysical perspective, we expect that if some new physics scenario changes the energy density of SM neutrinos from the $\Lambda$CDM value, there might be an associated change in the neutrino masses as well. 

We modify the Cosmic Linear Anisotropy Solving System (\texttt{CLASS})~\cite{Blas:2011rf,Lesgourgues:2011re,Lesgourgues:2011rh} code to account for different kinds of self-interactions and dark radiation isocurvature~\cite{Brinckmann:2022ajr,Ghosh:2021axu}. We have made this code publicly available.\footnote{The code is available at \href{https://github.com/subhajitghosh-phy/drisoCLASS}{https://github.com/subhajitghosh-phy/drisoCLASS}.} We then perform Markov chain Monte Carlo (MCMC) analysis with recent data sets to derive constraints on cosmological parameters.
In table~\ref{tab:summary}, we show a summary of constraints on new physics parameters with data from Planck 2018 temperature and polarization maps (PR3), baryon acoustic oscillation (BAO), and the South Pole Telescope (SPT-3G). We find no significant preference for physics beyond the $\Lambda$CDM model. However, data exhibit interesting interplay between massive neutrinos,  self-interacting DR as well as DR isocurvature. We summarize some main results here.
\begin{itemize}
    \item For adiabatic initial conditions, current data prefer the dominant contribution to $N_{\rm eff}$ to be nearly free-streaming and massless. For the cases with massless free-streaming or decoupling or recoupling DR and massive neutrinos, $N_{\rm dr}$ provides the dominant contribution to $N_{\rm eff}$. Therefore, $N_{\nu}$ is forced to be small, and the constraint on neutrino mass can be relaxed. This is of relevance given the recent DESI observations~\cite{DESI:2024mwx}. On the other hand, due to the strongly-coupled nature,  $N_{\rm dr}$ in the CDR case is constrained to be small. For detailed explanations, see section~\ref{subsec:AD}.
    \item For initial conditions with both adiabatic and DR isocurvature, data prefer a blue-tilted isocurvature power spectrum and larger $N_{\rm eff}$ is allowed. Similar to the adiabatic case, $N_{\rm dr}$ of CDR is small. However, $N_{\rm dr}$ and $N_{\nu}$ upper limits are comparable for other cases. For detailed explanations, see section~\ref{subsec:AD_ISO}.
    \item Regarding the $H_0$ tension, all secluded DR scenarios considered in this work cannot resolve the tension. The most improvement is seen for CDR with mixed adiabatic and DR isocurvature initial conditions. The Gaussian tension can be reduced to $3.5\sigma$ from 5.6$\sigma$. For detailed explanations, see section~\ref{sec:H0}.
\end{itemize}

\begin{table}[]
    \centering
Adiabatic Initial Condition
    \begin{tabular}{|c|c|c|c|c|}
    \hline
    \diagbox{Parameter}{Type} & Free-streaming & Coupled & Decoupling & Recoupling \\
    \hline
    \hline
      $\log_{10}\left(G_{\rm eff}~{\rm MeV}^2\right)$ & - & - & $<-4.10$ & - \\
      $\log_{10}\left(\lambda_{\rm eff}\right)$ & - & - & - & $-14.53^{+0.53}_{-0.39}$ \\
      $N_{\rm dr}$  &  $2.64^{+0.59}_{-0.068}$ & $0.281^{+0.098}_{-0.25}$ & $2.59^{+0.73}_{-0.074}$ & $2.62^{+0.63}_{-0.075}$ \\
      $N_\nu$   & $<0.235$ & $2.78\pm 0.21$ & $<0.315$ & $<0.276$ \\
      $N_{\rm eff}$ & $3.02 \pm 0.16$ & $3.06\pm 0.16$ & $3.05\pm 0.17$ & $3.04\pm 0.16$\\
      $N_\nu m_\nu (\rm{eV})$ & $<0.0761$ & $<0.0637$ & $<0.0796$ & $<0.0788$\\
      \hline
    \end{tabular}
    \\
    \vspace{2em}
    In the presence of Dark Radiation Isocurvature
    \begin{tabular}{|c|c|c|c|c|}
    \hline
    \diagbox{Parameter}{Type}  & Free-streaming & Coupled & Decoupling & Recoupling \\
    \hline
    \hline
      $10^{10} {\cal P}_{\rm{{\cal II}}}^{(1)}N_{\rm dr}^2$ & $<10.4$ & $<20.1$ & $<12.7$ & $<10.8$\\
      $10^{10} {\cal P}_{\rm{{\cal II}}}^{(2)}N_{\rm dr}^2$ & $37^{+7}_{-30}$ & $<215$ & $<58.6$ & $<53.6$\\
      $\log_{10}\left(G_{\rm eff}~{\rm MeV}^2\right)$ & - & - & $<-2.99$ & - \\
      $\log_{10}\left(\lambda_{\rm eff}\right)$ & - & - & - & $<-14.1$ \\
      $N_{\rm dr}$  &  $<2.83$ & $<0.309$ & $<2.40$ & $<2.79$\\
      $N_\nu$   & $<1.68$ & $2.94\pm 0.22$ & $<2.45$ & $<2.23$ \\
      $N_{\rm eff}$ & $3.10 \pm 0.17$ & $3.19^{+0.17}_{-0.19}$ & $3.11\pm 0.18$ & $3.14^{+0.17}_{-0.19}$\\
      $N_\nu m_\nu (\rm{eV})$ & $<0.0762$ & $<0.0719$ & $<0.0829$ & $<0.0829$\\
      \hline
    \end{tabular}
    \caption{Summary of constraints at $1\sigma$ or $68\%$ C.L. derived in this work, using the Planck+BAO+SPT data, on the new physics parameters. The interaction rates for decoupling and recoupling DR are denoted by $G_{\rm eff}$ and $\lambda_{\rm eff}$, defined in Eqs.~\eqref{eq:ave_rate_decoupling} and~\eqref{eq:ave_rate_recoupling}, respectively. The number of effective degrees of freedom in SM neutrinos and DR are denoted by $N_\nu$ and $N_{\rm dr}$, respectively, with their sum given by $N_{\rm eff}$ (Eq.~\eqref{eq:neff}). The energy in SM neutrinos, taking into their mass, is denoted by $N_\nu m_\nu$. Finally, the isocurvature power spectrum is parametrized at two scales $k_1 = 0.002 ~{\rm Mpc}^{-1}$ and $k_2 = 0.1 ~{\rm Mpc}^{-1}$ denoted by ${\cal P}_{\cal II}^{(1)}$ and ${\cal P}_{\cal II}^{(2)}$, respectively. To account for the degeneracy in ${\cal P_{II}}$ and $\Ndr$, we provide the constraint on the combination ${\cal P_{II}}\Ndr^2$.}
    \label{tab:summary}
\end{table}

This paper is organized as follows. We discuss the details of different DR scenarios in section~\ref{sec:DR_scenarios}. We then study the physical effects of these DR properties on the CMB observables in section~\ref{sec:observables}. We list the datasets used in our analysis in section~\ref{sec:datasets} and present results from MCMC runs in section~\ref{sec:MCMC_results}. In section~\ref{sec:H0}, we show how these DR scenarios fare in the context of the $H_0$ tension. We conclude in section~\ref{sec:conclusion}.

\section{Dark Radiation Scenarios}
\label{sec:DR_scenarios}

The total radiation density that decouples from the photon and baryon bath, including neutrinos and DR, is typically parametrized by $N_{\rm eff}$, defined as
\es{eq:neff}{
N_\eff \equiv {8\over 7}\left( \frac{11}{4} \right)^{4/3}\frac{\bar\rho_\nu + \bar\rho_\textrm{dr} }{\bar\rho_\gamma },
}
where $\bar\rho_\gamma$, $\bar\rho_\nu$, and $\bar\rho_{{\rm dr}}$ are the background energy densities of photons,  neutrinos, and  DR, respectively. We further define $N_\nu$ and $N_{\rm dr}$ to denote the contribution to $N_{\rm eff}$ from SM neutrinos and DR separately, with $N_{\rm eff}=N_\nu+N_{\rm dr}$. The standard $\Lambda$CDM model predicts $N_\nu$ to be a fixed number: $N_\nu^{\Lambda\textrm{CDM}}=3.044$~\cite{Froustey:2020mcq,Bennett:2020zkv,Akita:2020szl}. In this work, we consider new physics dynamics that can change $N_\nu$ and generate sizable $N_{\rm dr}$. For example, this can occur in scenarios where new heavy particles decay into the SM bath and dark radiation after neutrino decoupling. Alternatively, converting  neutrino energy density to DR prior to the formation of CMB can have the same effects. In this case, DR is still secluded at the time of CMB decoupling. Therefore, we allow both $N_\nu$ and $N_{\rm dr}$ to vary in our study. 

We will use the synchronous gauge to describe the time evolution of perturbations.
In this gauge the perturbations around a flat Friedmann-Robertson-Walker metric can be written as 
\es{}{
\D s^2=a^2(\tau)(-\D \tau^2+(\delta_{ij}+H_{ij})\D x^i \D x^j),
}
where $\tau$ is the conformal time and $a(\tau)$ is the scale factor. The metric perturbation $H_{ij}$ in Fourier space is given by,
\es{eq:synch}{
    H_{ij}(\vec{x}, \tau) = \int \D^3 k e^{i\vec{k}\cdot \vec{x}} H_{ij}(\vec{k},\tau),~~
H_{ij}(\mathbf{k},\tau)=\hat { k}_i\hat  { k}_j h(\mathbf k,\tau) +\left(\hat {k}_i\hat{k}_j -\frac{1}{3}\delta_{ij}\right)6\eta(\mathbf k,\tau),
}
where $h $ and $\eta$ denote the trace and traceless longitudinal part of $H_{ij}$ respectively. 
Here $\mathbf{k}\equiv k \hat {\mathbf k} $ is Fourier conjugate to comoving position $\mathbf x$.

The fluctuations in DR can be described as a perturbation to the phase space distribution of DR $f(\mathbf{q},\mathbf{k},\tau)$
\bea\label{eq:f_expansion}
f(\mathbf{q},\mathbf{k},\tau)\equiv \bar f(q,\tau) (1+\Psi(\mathbf{q},\mathbf{k},\tau)),
\eea
where $\bar f(q,\tau)$ is the averaged phase space distribution and $\mathbf{q}\equiv q \hat{\mathbf {q}}$ is the comoving momentum of DR, which is related to the physical momentum $\mathbf p$ as $\mathbf{q}=a \mathbf{p}$. In this work, we assume that all kinds of DR in our analysis have nearly thermal distributions,\footnote{This condition can be easily satisfied for DR that has strong self-interactions like fluid or decoupling scenarios. For weak self-interactions (recoupling or free-streaming scenarios), we assume DR is a thermal relic, like SM neutrinos, that inherits thermal distribution from other interactions before decoupling.} which means the averaged phase space distribution is set to the equilibrium distribution, 
\beq
\bar f(q,\tau)\approx \bar{f}^{\rm eq}(q) \approx N e^{-q/T_{D,0}},
\eeq 
where $T_{D,0}$ is the temperature of DR today, and we have used the Maxwell-Boltzmann distribution as an approximation. The normalization factor $N$ is chosen to match the \textit{energy} density between the Maxwell-Boltzmann and Bose-Einstein/Fermi-Dirac distribution: $\int dq\, q^3 \bar{f}^{\rm eq}(q)=\int dq\, q^3 f^{\rm BE/FD}(q)$. Therefore, we get  $N= \pi^4/90$ for bosons and $N=7 \pi^4/720$ for fermions.

As mentioned above, the non-trivial time evolution of $f(\mathbf{q},\mathbf{k},\tau)$ appears in the perturbation $\Psi(\mathbf{q},\mathbf{k},\tau)$ defined in Eq.~(\ref{eq:f_expansion}). The Boltzmann equation for $\Psi $ is given as:
\bea\label{eq:Boltzmann_eq_general}
\dot{\Psi}+i k P_1( \hat {\mathbf k} \cdot \hat {\mathbf q})\Psi+\frac{\D \ln \bar f}{\D \ln q}\left[-\frac{\dot h}{6}-\frac{P_2( \hat {\mathbf k} \cdot \hat {\mathbf q})}{3}(\dot h+6\dot \eta)\right]=C[f],
\eea
where $P_\ell$ are the Legendre polynomials and an overdot $\, \dot{\langle ~\rangle} \,$ denotes a derivative with respect to $\tau$. $C[f]$ is the collision term that accounts for DR interactions. The full expression of $C[f]$ was derived in \cite{Oldengott:2014qra,Oldengott:2017fhy}. 
Since we deal with purely relativistic particles, it is convenient to use $F$ instead of $\Psi$ in the analysis,
\bea\label{eq:Fell}
F(\mathbf k, \hat{\mathbf q},\tau)\equiv\frac{\int \D q q^3\bar  f\Psi}{\int \D q q^3\bar  f}\equiv \sum_{\ell=0}^{\infty}(-i)^\ell (2\ell+1) F_{\ell}(k,\tau)P_\ell( \hat {\mathbf k} \cdot \hat {\mathbf q}) .
\eea
The Boltzmann hierarchy depends on the exact form of the collision term. We will classify DR with the following categories.

\subsection{Free-streaming and Coupled DR}
For free-streaming DR (FDR), the self-interaction is negligible compared to the Hubble rate. Therefore, we can set the collision term in Eq.~(\ref{eq:Boltzmann_eq_general}) to zero. Then the Boltzmann hierarchy  for $ F_{ \ell}$ is given as
  \bea\label{eq:Boltzmann_hierarchy_fs}
&&\dot{\delta}=-\frac{4}{3}\theta-\frac{2}{3}\dot h,\nn
&&\dot{\theta}=\frac{k^2}{4}\delta-k^2 \sigma,\nn
&&\dot{\sigma}=\frac{4}{15}\theta -\frac{3k}{10}F_{3}+\frac{2}{15}\dot h+\frac{4}{5}\dot \eta,\nn
&&\dot{F}_{ \ell}=\frac{k}{2\ell+1}[\ell F_{\ell-1}-(\ell +1)F_{\ell+1}], ~~~~(\ell\geq3).
\eea
Here, we have used the density contrast $\delta \equiv F_0$, the velocity perturbation $\theta \equiv (3k/4) F_1$, and the anisotropic stress $\sigma \equiv F_2/2$.
This is the same Boltzmann hierarchy for free-streaming massless neutrinos. Just as neutrinos, free-streaming DR can generate anisotropic stress $\sigma$ and higher moments of perturbations. Their perturbations propagate at the speed of light, faster than those of photon and baryon bath. Therefore, photon perturbations experience gravitational pull from free-streaming radiation, resulting in a phase shift in the CMB anisotropy angular power spectrum~\cite{Bashinsky:2003tk, Baumann:2015rya}. In addition to the phase shift, the presence of additional radiation density increases the Hubble expansion parameter, leading to a stronger damping in the CMB power spectrum. 

If DR is strongly-coupled, meaning the self-interaction rate is always greater than the Hubble expansion rate, DR behaves like a coupled fluid. In this case, the anisotropic stress and higher moments of DR perturbations $F_\ell$ ($\ell\ge 2$) are dynamically driven to zero. Therefore, we can simplify the Boltzmann hierarchy for coupled DR (CDR) as 

\bea\label{eq:Boltzmann_hierarchy_coupled}
&&\dot{\delta}=-\frac{4}{3}\theta-\frac{2}{3}\dot h,\nn
&&\dot{\theta}=\frac{k^2}{4}\delta.
\eea
CDR behaves similarly to photons, and therefore does not generate much phase shift in the CMB angular power spectrum as FDR, as discussed further in the next section. Moreover, compared to the FDR case, CDR leads to less damping in the CMB power spectrum due to a boost when modes reenter the horizon (see section~\ref{sec:observables}). Generally, CMB data allow for a larger $N_{\rm dr}$ in CDR than in FDR (see, e.g.,~\cite{Blinov:2020hmc}).

\subsection{Recoupling and Decoupling DR}
\label{sec:intDR}
Dark radiation may transition between free-streaming and fluid-like behavior depending on how its self-interaction rate scales relative to the Hubble rate. If this transition occurs either long before or well after matter-radiation equality, the CMB data are influenced by only one of these two stages, i.e., purely free-streaming or purely fluid-like behavior. However, if the transition takes place near equality, the effects of both stages must be considered. 
We refer to the DR that was fluid in the early Universe and later became free-streaming as decoupling DR, while recoupling DR describes the opposite case.

In this study, we use the relaxation time approximation to analyze decoupling and recoupling DR, following the treatment of \cite{Oldengott:2014qra,Oldengott:2017fhy,Brinckmann:2022ajr}. Under this approximation, the leading contribution to the collision term for each perturbation is linear in this perturbation. Then the Boltzmann hierarchy becomes
\bea\label{eq:Boltzmann_hierarchy_decoupling}
&&\dot{\delta}=-\frac{4}{3}\theta-\frac{2}{3}\dot h,\nn
&&\dot{\theta}=\frac{k^2}{4}\delta-k^2 \sigma,\nn
&&\dot{\sigma}=\frac{4}{15}\theta -\frac{3k}{10}F_{3}+\frac{2}{15}\dot h+\frac{4}{5}\dot \eta-a \effrate \sigma,\nn
&&\dot{F}_{ \ell}=\frac{k}{2\ell+1}[\ell F_{\ell-1}-(\ell +1)F_{\ell+1}]- a \effrate {F}_{\ell}, ~~~~\ell\geq3,
\eea
where $\effrate$ is a model-dependent effective interaction rate. In principle, $\effrate$ is slightly different for different $\ell$, but we take the one for $\ell=2$ since the effect of collision terms is dominated by the second moment $\sigma$. We note that the effective rate for $\delta$ and $\theta$ is vanishing due to energy and momentum conservation, respectively. If $\effrate \propto T^n$ with $n>2$, $\effrate$ decreases faster than the Hubble rate, which scales as $T^2$ in RD, leading to DR decoupling. In contrast, if $n<2$, the DR will recouple at a lower temperature for RD.

In this work, we consider one example for each case. For decoupling DR, we examine the self-interactions of a Majorana fermion $\chi$ (DR) mediated by a heavy particle $\phi$. This interaction corresponds to the neutrino self-interaction through a heavy Majoron. The Lagrangian relevant for this scenario is
\beq\label{eq:Ldec}
-\mathcal{L} \supset \frac{1}{2}m_\phi^2 \phi^2 +\frac{1}{2}g_\phi \phi \bar\chi\chi \approx \frac{1}{8} G_\phi \bar{\chi}\chi\bar{\chi}\chi \, ,
\eeq
where the approximation in Eq.~\eqref{eq:Ldec} with $G_\phi \equiv g_\phi^2/m_\phi^2$ holds once $\phi$ is integrated out at low temperatures. The effective interaction rate is given as\footnote{For the Dirac fermion case with $-\mathcal{L} \supset g_\phi \phi\bar\chi\chi$, we can simply replace $g_\phi$ with $2g_\phi$, or equivalently $G_\phi$ with $4 G_\phi$.}
\beq\label{eq:ave_rate_decoupling}
\effrate = \alpha_{2,\textrm{dec}} \avgrate =\frac{7\pi \alpha_{2,\textrm{dec}}}{576} G_\phi^2T_{D}^5 = \Geff ^2T_{D}^5,
\eeq
where $T_{D}$ is the temperature of DR, which we take to be the SM neutrino temperature $T_D=T_\nu$ for simplicity. Other cases with different DR temperatures can be obtained by rescaling the interaction rates accordingly. The quantity $\alpha_{2,\textrm{dec}}=1.39$ is the relaxation time coefficient for $\ell=2$~\cite{Brinckmann:2022ajr}, which matches the collision terms in the Boltzmann hierarchy with the thermal averaged rate $\avgrate$ (for details, see appendix~\ref{app:boltzmann_eq}). This relation gives $\Geff=0.23~G_\phi$.
Note that $\effrate \propto a^{-5}$ blows up for very small $a$ (early times). This limit corresponds to the radiation behaving as a coupled fluid. To avoid numerical issues in the strongly-coupled regime, we set a cutoff on $a\effrate$: $a\effrate=\textrm {Min}[a G_{\rm eff}^2 T_{D}^5,\Lambda]$, with $\Lambda$ chosen to be $100/\textrm{Mpc}$ in \texttt{drisoCLASS} code. We have checked numerically that the observables are not sensitive to the precise value of $\Lambda$ for $\Lambda \ge 100/\textrm{Mpc}$.

For an example of recoupling DR, we consider a light scalar $\phi$ with the $\phi^4$ interaction, where the Lagrangian is
\beq
-\mathcal{L} \supset \frac{\lambda_\phi}{4!} \phi^4 \, ,
\eeq
where $\lambda_\phi$ denotes the coupling constant.
The interaction rate for this case is given by
\beq\label{eq:ave_rate_recoupling}
\effrate = \alpha_{2,\textrm{rec}} \avgrate = \frac{\pi \alpha_{2,\textrm{rec}}}{23040} \lambda_\phi^2 T_{D} = \lambda_\textrm{eff}^2 T_D\, ,
\eeq
where we set $T_D=T_\nu$ and the relaxation time coefficients as 
$\alpha_{2,\textrm{rec}}=0.188$, which gives $\lambda_\textrm{eff} = 5.1\times10^{-3}\lambda_\phi$ (for details, see appendix~\ref{app:boltzmann_eq}).

For both cases, we define the redshifts of decoupling and recoupling as
\begin{eqnarray}\label{eq:def_z_dec_rec}
   \effrate(z) = H(z)\Big|_{z=z_{\rm dec/rec}}.
\end{eqnarray}

\subsection{Initial Conditions}
Before discussing the dynamics of cosmological perturbations in detail, we summarize the different initial conditions (IC).
Adiabatic IC have been widely studied; a summary can be found in Ref.~\cite{Ma:1995ey}.
Isocurvature IC have also been studied for baryons, CDM, and neutrinos~\cite{Bucher:1999re}, and for DR~\cite{Ghosh:2021axu}.
Here, we provide a brief overview of how to derive these IC for completeness.

\paragraph{Equations of Motion.} The perturbed Einstein equations in the synchronous gauge are given by~\cite{Ma:1995ey},
\es{eq:ee_pert}{
k^2\eta - {1\over 2}{\cal H}\dot{h} &= -{3\over 2}{\cal H}^2 \sum_i f_i \delta_i,\\
k^2\dot{\eta} &= {3\over 2}{\cal H}^2 \sum_i (1+w_i)f_i  \theta_i,\\
\ddot{h}+2 {\cal H} \dot{h} -2k^2\eta &= -9{\cal H}^2 \sum_i c_{s,i}^2 f_i \delta_i,\\
\ddot{h} + 6 \ddot{\eta} + 2{\cal H}(\dot{h}+6\dot{\eta})-2k^2\eta &= -9{\cal H}^2 \sum_i(1+w_i)f_i\sigma_i,
}
where the index $i$ goes over all the species (photons: $\gamma$, neutrinos: $\nu$, baryons: $b$, DM: $c$, DR: ${\rm dr}$, and total radiation: $r$).
For species $i$, the pressure is related to the energy density as $\bar{p}_i = w_i \bar{\rho}_i$; the fractional abundance $f_i = \bar{\rho}_i / \sum_i \bar{\rho}_i$, the speed of sound $c_{s,i}^2 = \partial \bar{p}_i /\partial \bar{\rho}_i$, the density contrast $\delta_i = \delta\rho_i/\bar{\rho}_i$, the velocity perturbation $\theta_i$, and the anisotropic stress $\sigma_i$.
These equations need to be supplemented by the dynamics governing the time evolution of each species.
At very early times, much before recombination, baryons are tightly coupled to photons, and that forces their velocities to be equal $\theta_b = \theta_\gamma$.
Furthermore, in the synchronous gauge we can choose $\theta_c=0$.
The baryon density and the CDM density then evolve via,
\es{eq:bar_cdm}{
\dot{\delta}_b + \theta_b + {1\over 2}\dot{h} = 0,\\
    \dot{\delta}_c + {1\over 2}\dot{h} = 0.
}
The early time evolution of photons (similar to CDR), neutrinos (similar to FDR), and DR can be obtained from Eqs.~\eqref{eq:Boltzmann_hierarchy_fs} and~\eqref{eq:Boltzmann_hierarchy_decoupling}, and to derive initial conditions, we only keep $\delta,\theta,\sigma$ because $F_{\ell\geq 3}$ are negligible at early times.
Since photons and CDR are tightly coupled, we can further ignore the anisotropic stress term. 
We note that although decoupling DR is strongly coupled at early times due to strong self-interactions, we set its IC as FDR.
The non-zero initial value of $\sigma$ is rapidly driven to zero due to large $\Gamma_{\rm eff}$ at early times, and thus the initial choice of $\sigma$ does not impact the cosmological observables.
This treatment of decoupling DR is also consistent with evolution functions in Eq.~\eqref{eq:Boltzmann_hierarchy_decoupling} as shown in Ref.~\cite{Brinckmann:2022ajr}.  
Therefore, for neutrinos, FDR, decoupling, and recoupling DR, the Boltzmann hierarchy relevant for deriving initial conditions is
\begin{eqnarray}\label{eq:nu_fdr_boltzmann}
   \dot{\delta}_{\nu,\textrm{dr}} +{4\over 3}\theta_{\nu,\textrm{dr}} + {2 \over 3}\dot{h}=0,  \\
    \dot{\theta}_{\nu,\textrm{dr}} -k^2\left({1\over 4}\delta_{\nu,\textrm{dr}} -\sigma_{\nu,\textrm{dr}}\right)=0,\\
    \dot{\sigma}_{\nu, \textrm{dr}} - {4\over 15}\theta_{\nu, \textrm{dr}} - {2\over 15}\dot{h} - {4\over 5}\dot{\eta} = 0.
\end{eqnarray}
For photons and CDR, it gives
 \begin{eqnarray}\label{eq:gamma_cdr_boltzmann}
   \dot{\delta}_{\gamma, \textrm{dr}} +{4\over 3}\theta_{\gamma, \textrm{dr}} + {2 \over 3}\dot{h}=0, \\
    \dot{\theta}_{\gamma,\textrm{dr}} -\frac{k^2}{4}\delta_{\gamma,\textrm{dr}}=0.
\end{eqnarray}
We also note that in the early Universe, when radiation and matter dominate, the conformal Hubble rate can be expressed as,
\es{}{
{\cal H} = {1\over \tau}{1+\omega \tau/2 \over 1+\omega\tau/4},
}
where $\omega = H_0\Omega_m/\Omega_r^{1/2}$ and $\Omega_{m/r}\equiv \bar\rho_{m/r,0}/\rho_{c,0}$ is the fractional density of matter or radiation today with respect to the present day critical density $\rho_{c,0}$.
We set IC when $\omega \tau \rightarrow 0$.
However, to track the IC accurately, we need to keep certain quantities to the linear order in $\omega \tau$.
To that end, deep during radiation domination, we derive $\rho_m = \omega\tau \rho_{\rm tot} (1+\omega\tau/4)/(1+\omega\tau/2)^2\approx \omega\tau\rho_{\rm tot}$.

\paragraph{Gauge Invariant Perturbations.} To express IC in a gauge invariant way, we start with the first-order scalar fluctuations of the metric (for a review see~\cite{Malik:2008im})
\es{eq:gen_metric}{
    \D s^2 = a^2\left((-1-2\phi)\D\eta^2 + 2 B_{,i} \D \eta \D x^i + (\delta_{ij} -2\psi \delta_{ij}+2E_{,ij})\D x^i \D x^j\right),
}
where $B_{,i} = \partial_i B$ and $E_{,ij} = \partial_i\partial_j E$.
The gauge invariant curvature perturbation is given by,
\es{}{
    \zeta = -\psi - H {\delta\rho \over \dot{\bar{\rho}}}.
}
A similar quantity can be defined to denote gauge invariant fluctuations in a species $i=\gamma,\nu,b,c,{\rm dr}$,
\es{}{
    \zeta_i = -\psi - H {\delta\rho_i \over \dot{\bar{\rho}}}.
}
The isocurvature perturbation in species $i$ with respect to species $j$ is then defined as,
\es{}{
    S_{ij} = 3(\zeta_i-\zeta_j).
}
It is convenient to decompose the set of IC for primordial perturbations into adiabatic ($S_{ij}=0$) and isocurvature ($\zeta = 0$) ones.
A generic initial condition will then be a linear combination of the two.

\paragraph{Adiabatic Initial Conditions.} To determine the adiabatic IC, we require $S_{ir}\equiv 3(\zeta_i - \zeta_r)=0$ for all species, where $\zeta_i$ and $\zeta_r$ are gauge invariant perturbations in species $i$ and total radiation $r$, respectively, with $\zeta_r = \sum_{i=\gamma,\nu,{\rm dr}} R_i \zeta_i$ where $R_i = \bar{\rho}_i/(\bar{\rho}_\gamma+\bar{\rho}_\nu+\bar{\rho}_{\rm dr})$.
Demanding $S_{ir}=0$ forces,
\es{}{
\delta_\gamma = \delta_{\rm dr} = \delta_\nu = {4\over 3}\delta_b = {4\over 3}\delta_c.
}
Furthermore, for the fastest growing adiabatic mode, the solution is $\theta_i = 0$ at the order $\tau^0$~\cite{Ratra:1988bz, Ma:1995ey}.
Using these conditions, we can derive the IC for all the perturbations with FDR, decoupling and recoupling DR:
\es{eq:adia_ic}{
\delta_\gamma &= \delta_\nu = \delta_{\rm dr} =  (4/3)\delta_b = (4/3)\delta_c = -(1/3)(k\tau)^2,\\
\theta_\gamma &= \theta_b = -(1/36)k (k \tau)^3,\\
\theta_\nu &= \theta_{\rm dr} = -(1/36)(23+4 R_\nu+4R_{\rm dr})/(15+4 R_\nu + 4 R_{\rm dr})k (k \tau)^3,\\
\sigma_\nu &= \sigma_{\rm dr} = (2/3)(k\tau)^2/(15+4 R_\nu + 4R_{\rm dr}),\\
h &= {1\over 2}(k\tau)^2,\\
\eta &= 1 - (5+4 R_\nu + 4 R_{\rm dr})/(12(15+4 R_\nu + 4 R_{\rm dr}))(k\tau)^2.
}
On the other hand, for CDR, the IC are
\es{eq:iso_ic}{
\delta_\gamma &= \delta_\nu = \delta_{\rm dr} =  (4/3)\delta_b = (4/3)\delta_c = -(1/3)(k\tau)^2,\\
\theta_\gamma &= \theta_b =\theta_{\rm dr} = -(1/36)k (k \tau)^3,\\
\theta_\nu & = -(1/36)(23+4 R_\nu)/(15+4 R_\nu)k (k \tau)^3,\\
\sigma_\nu &= (2/3)(k\tau)^2/(15+4 R_\nu),\\
h &= {1\over 2}(k\tau)^2,\\ \eta &= 1 - (5+4 R_\nu)/(12(15+4 R_\nu))(k\tau)^2.
}
The main difference between the two originates from vanishing $\sigma_{\rm dr}$ for CDR.

\paragraph{Isocurvature Initial Conditions.}
In this work, we consider the initial condition involving isocurvature being DR isocurvature density mode (DRID), which is similar to the standard neutrino isocurvature density mode~\cite{Bucher:1999re}.
To fix isocurvature IC for FDR and CDR, we assume baryons and CDM perturbations are adiabatic and set $S_{br}=S_{cr}=0$.
Furthermore, we also demand the total curvature perturbation $\zeta \rightarrow 0$ as $\tau \rightarrow 0$.
In the synchronous gauge $\zeta$ can be written as (see appendix~\ref{sec:app_gauge} for a derivation), $\zeta = -\eta - {\cal H}\delta\rho/\rho'$.
Therefore, by demanding $\eta\rightarrow 0$ and  $\delta\rho\rightarrow 0$ as conformal time $\tau\rightarrow 0$, we select isocurvature IC.
Since these IC are fixed deep during radiation domination, they also force $\delta \rho_r\rightarrow 0$ as $\tau\rightarrow 0$.
To determine the initial condition for neutrino perturbations, we fix $S_{\nu \gamma}=0$.\footnote{One can alternately choose $S_{\nu r}=0$ which can be realized in some particular models.}
Normalizing $\delta_{\rm dr}\rightarrow 1$ as $\tau\rightarrow 0$, the above conditions lead to
$\delta_\gamma = \delta_\nu = -\rdr/(1-\rdr)$ as $\tau\rightarrow 0$.
The early-time asymptotic behavior for all the relevant cosmological perturbations are summarized in table~\ref{tab:iso_ic}, following Ref.~\cite{Ghosh:2021axu}. Here we keep terms up to $O((k\tau)^2)$. For terms that are zero up to this order, we show the leading non-vanishing term.
\begin{table}[]
    \centering
    \begin{tabular}{|c|c|c|}
    \hline
       & FDR & CDR \\
       \hline
       \hline
    $\delta_\gamma$  & ${\rdr \over 1-\rdr}\left(-1 + {(k\tau)^2\over 6}\right)$ & Same as FDR \\
    $\theta_\gamma$ & $-{\rdr \over 1-\rdr}{k^2\tau \over 4}$ & Same as FDR \\
    $\delta_{\rm dr}$ & $1- {(k\tau)^2\over 6}$ & Same as FDR \\
    $\theta_{\rm dr}$ & ${k^2\tau \over 4}$ & Same as FDR \\
    $\sigma_{\rm dr}$ & ${15 -15 \rdr +4 R_\nu \over (1-\rdr )(15+4\rdr+4 R_\nu)}{(k\tau)^2 \over 30}$ & 0 \\
    $\delta_b$ & ${\rdr \over 1-\rdr}{(k\tau)^2 \over 8}$ & Same as FDR \\
    $\delta_c$ & $-{\rdr R_b \over 1-\rdr}{k^2\omega\tau^3 \over 80}$ & Same as FDR \\
    $\delta_\nu$ & ${\rdr \over 1-\rdr}\left(-1 + {(k\tau)^2\over 6}\right)$ & Same as FDR \\
    $\theta_\nu$ & $-{\rdr \over 1-\rdr}{k^2\tau \over 4}$ & Same as FDR \\
    $\sigma_\nu$ & $-{19\rdr \over (1-\rdr)(15+4\rdr+4 R_\nu)}{k^2\tau^2 \over 30}$ & $-{\rdr \over (1-\rdr)(15+4 R_\nu)}{k^2\tau^2 \over 2}$ \\
    $h$ & ${\rdr R_b \over 1-\rdr}{k^2\omega\tau^3 \over 40}$ & Same as FDR \\
    $\eta$ & $-{\rdr-\rdr^2 - \rdr R_\nu \over (1-\rdr)(15+4\rdr+4 R_\nu)}{k^2\tau^2 \over 6}$ & ${\rdr R_\nu \over (1-\rdr)(15+4 R_\nu)}{k^2\tau^2 \over 6}$ \\
    \hline
    \end{tabular} 
    \caption{
    Isocurvature initial conditions of cosmological perturbations for FDR and CDR. Decoupling and recoupling DR obey the same IC as FDR. Here $R_{b(c)}=\Omega_{b(c)}/\Omega_m$ is the fraction of matter in the form of baryons (CDM). 
    }
    \label{tab:iso_ic}
\end{table}

\paragraph{Primordial Power Spectrum.} The physical observables depend not only on the time evolution of perturbations, but also on their primordial spatial distribution, often encoded in the primordial power spectrum. In this study, we assume there is no modification to the adiabatic mode. Therefore, the adiabatic primordial power spectrum follows the standard form as in the $\Lambda$CDM model:
\begin{eqnarray}
    \mathcal{P}_{\cal R\cal R}(k)=A_s \left(k \over k_\ast\right)^{n_s-1},
\end{eqnarray}
where $A_s$ is the amplitude and $n_s$ is the spectral index of the adiabatic power spectrum. We take pivot scale $k_\ast=0.05~\textrm{Mpc}^{-1}$ consistent with the Planck collaboration~\cite{Planck:2018vyg}. 

For the isocurvature power spectrum, we assume it has a similar form that follows a simple power law, which is the natural consequence of axion or curvaton models~\cite{Planck:2018jri}. We then parametrize the DR isocurvature spectrum as
\begin{eqnarray}\label{eq:P_iso}
    \mathcal{P}_{\cal I\cal I}(k)=f_{\rm drid}^2A_s \left(k \over k_\ast\right)^{n_{\rm drid}-1},
\end{eqnarray}
where $f_{\rm drid}^2A_s$ is the amplitude and $n_{\rm drid}$ is the spectral index of DR isocurvature power spectrum. In the later MCMC analysis, we will treat $f_{\rm drid}$ and $n_{\rm drid}$ as additional free parameters. Therefore, the DRID spectrum does not have to be nearly scale invariant as the adiabatic spectrum.

\section{Physical Effects on the CMB Observables}\label{sec:observables}
In this section, we will discuss the impact of various DR properties mentioned in section~\ref{sec:DR_scenarios}, including different kinds of self-interactions and IC, on the CMB observables. Certain aspects of these effects of DR have been studied in the literature: free-streaming and self-interacting DR (decoupling, recoupling, and coupled DR) with adiabatic IC were studied in Refs.~\cite{Bashinsky:2003tk, Cyr-Racine:2013jua, Baumann:2015rya, Choi:2018gho, Archidiacono:2013dua,Kreisch:2019yzn,Brinckmann:2022ajr}; FDR and CDR with isocurvature IC were presented in Ref.~\cite{Ghosh:2021axu}. This work completes the analysis by including isocurvature IC for decoupling and recoupling DR. 
To illustrate the physical effects of various parameters, we focus on the temperature angular power spectrum ($C_{\ell}^{\rm TT}$) and the lensing power spectrum ($C_{L}^{\phi\phi}$). 
Before proceeding, it is useful to recall some relevant quantities:
\es{}{
&{\rm sound~horizon~at~decoupling}: r_s = \int_0^{a_\star}{\D a \over a^2 H}c_s,\\
&{\rm diffusion~scale~at~decoupling}: r_d^2 = \pi^2\int_0^{a_\star}{\D a \over 6(1+R)n_e \sigma_T a^3 H}\left({R^2 \over 1+R}+{16\over 15}\right),\\
&{\rm angular~diameter~distance~to~}z_\star: D_M = \int_{a_\star}^1 {\D a \over a^2 H}.
}
Here, we have denoted the free electron density by $n_e$, Thomson scattering cross section by $\sigma_T$, the scale factor at CMB decoupling by $a_\star$, the associated redshift by $z_\star$, and $R=3\bar{\rho}_b/(4\bar{\rho}_\gamma)$.
The angular scale corresponding to $r_s$ and $r_d$ are denoted by $\theta_s=r_s/D_M$ and $\theta_d=r_d/D_M$, respectively, with $\theta_s$ measured at an exquisite precision.
The two angular scales are related via $\theta_d = (r_d/r_s)\theta_s$.

\begin{figure}
    \centering
    \includegraphics[width=0.47\linewidth]{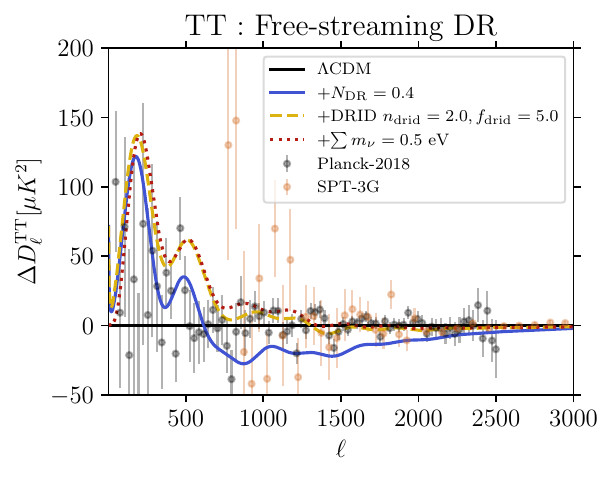}
    \includegraphics[width=0.47\linewidth]{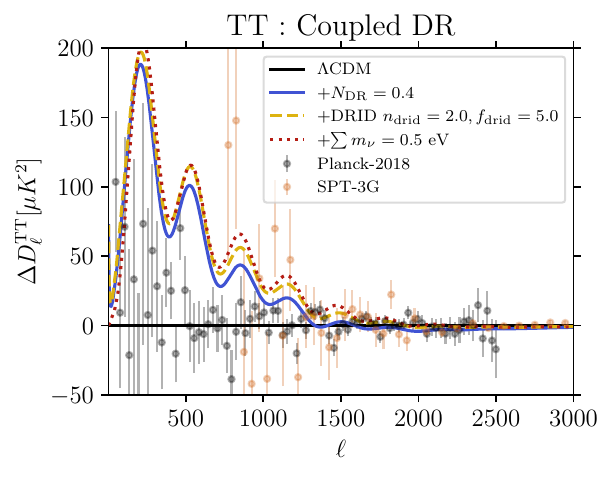}
    \includegraphics[width=0.47\linewidth]{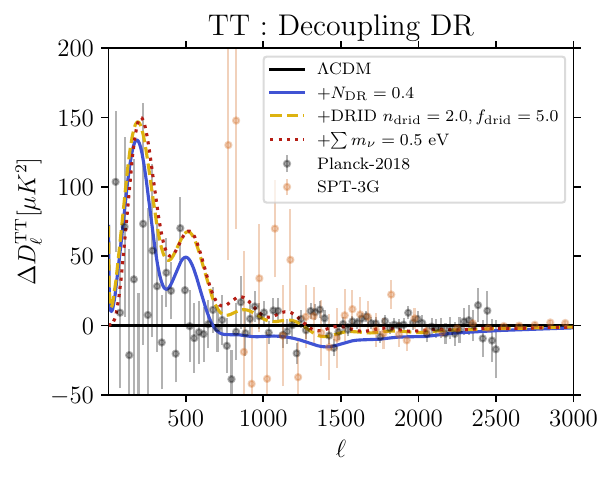}
    \includegraphics[width=0.47\linewidth]{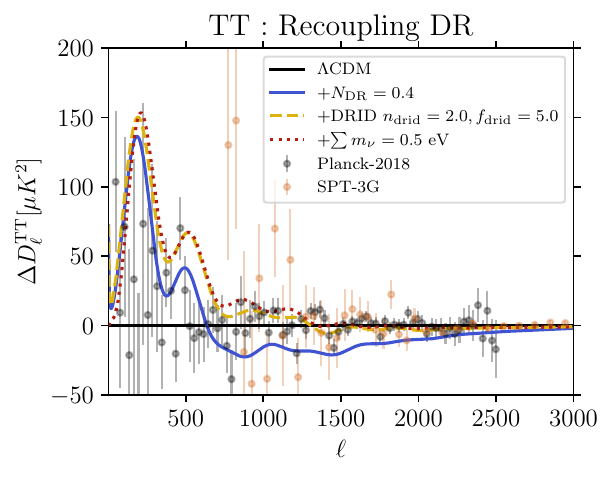}
    
    \caption{The CMB temperature anisotropy power spectrum ($D^{\rm TT}_\ell\equiv T_0^2\ell(\ell+1)C_\ell^{\rm TT}/(2\pi)$) for different cases of DR scenarios: FDR, CDR, decoupling and recoupling DR. For each type of DR, we show the effects of adiabatic and  isocurvature initial conditions, as well as massive neutrinos, comparing to the $\Lambda\rm CDM$ case. We also show data points from Planck 2018~\cite{Planck:2019nip} and SPT-3G~\cite{SPT-3G:2022hvq}.}
    \label{fig:TT}
\end{figure}

\paragraph{Adiabatic Initial Conditions.} We first discuss cases with adiabatic IC. 
Adding extra radiation via FDR increases $H$ and decreases both $r_s$ and $r_d$, with their ratio $r_d/r_s \propto \sqrt{H}$ increasing.
Since the data fix $\theta_s$ with high precision from measurement of the acoustic peaks, this leads to an increase in $\theta_d$.
This implies Silk damping sets in at smaller values of $\ell$.
Another effect of adding radiation is a later onset of matter-radiation equality, which leads to enhancements of the first few peaks compared to $\Lambda$CDM, via the `radiation driving' effect~\cite{Hu:1995en}.
In particular, acoustic oscillations get boosted after mode re-entry during radiation domination due to gravitational potential decay.
On the other hand, modes re-entering during matter domination do not experience this boost.
Therefore, pushing the matter-radiation equality to later epochs implies more modes experience the boost, leading to enhancements of the first few peaks.
These features are manifest in the top left panel of figure~\ref{fig:TT}, where we have fixed $\theta_s$ to its $\Lambda$CDM value.\footnote{To have a better comparison with the experimental data, we show $D^{\rm TT}_\ell\equiv T_0^2\ell(\ell+1)C_\ell^{\rm TT}/(2\pi)$ in figure~\ref{fig:TT}, where $T_0$ is the averaged temperature of the CMB today~\cite{Planck:2018vyg}.} 
Adiabatic FDR also results in a phase shift (compared to the case where there in no free-streaming species) because the acoustic modes in the bath experience gravitational pull from FDR perturbations that travel at the speed of light ($c = 1$)~(see e.g.,~\cite{Bashinsky:2003tk, Baumann:2015rya, Blinov:2020hmc,Montefalcone:2025unv}). 
This total phase shift $(\theta)$ can be estimated in terms of the fractional energy density in free-streaming radiation, which includes SM neutrinos, compared to total radiation,
\es{eq:ps}{
\theta \simeq 0.191 \pi { \bar{\rho}_\nu + \bar{\rho}_{\rm fdr} \over \bar{\rho}_r}.
}

\begin{figure}
    \centering
    \includegraphics[width=0.45\linewidth]{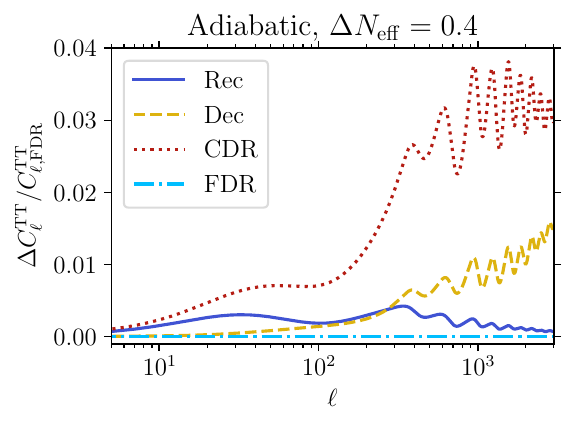}
    \includegraphics[width=0.45\linewidth]{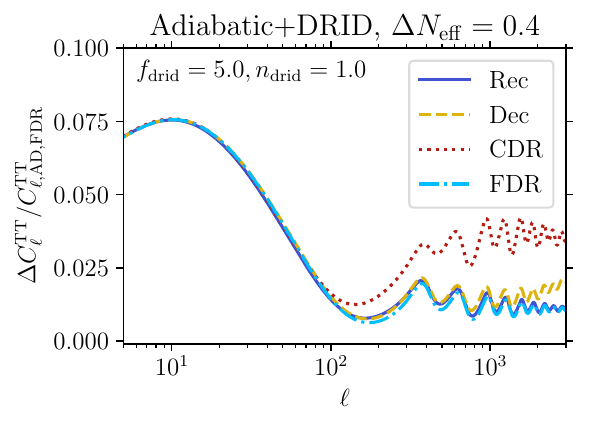}
    \includegraphics[width=0.45\linewidth]{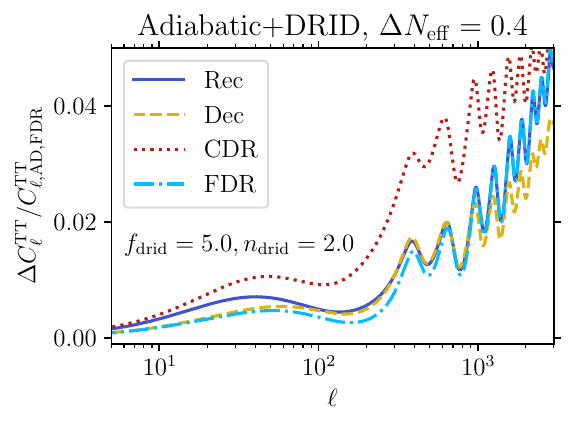}
    \includegraphics[width=0.45\linewidth]{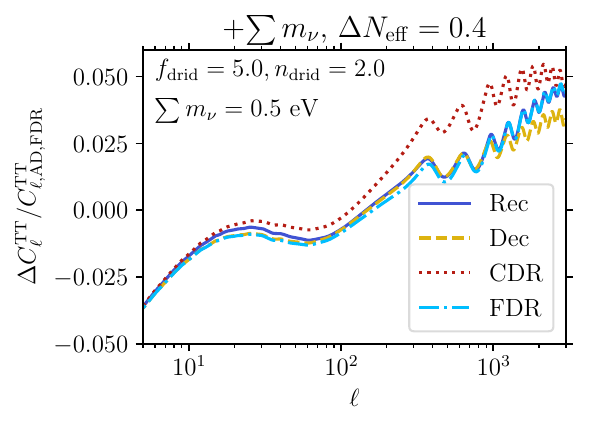}
    
    \caption{The fractional difference of $C^{\rm TT}_\ell$ with respect to the adiabatic FDR case for different cases of DR scenarios with the same $\Delta N_{\rm eff}=0.4$. Therefore, the FDR line in the top left panel is flat. We show the effects of including isocurvature IC (top right and bottm left), as well as massive neutrinos (bottom right).}
    \label{fig:TT_frac}
\end{figure}

CDR, similar to FDR, enhances the first few peaks as well.
However, the CMB anisotropies increase in size compared to FDR.
One way to see this is to look at the initial condition for the Sachs-Wolfe term $(1/4)\delta_\gamma+\psi$ which goes as $5/(15+4R_\nu)$ for CDR and $5/(15+4R_\nu+4\rdr)$ for FDR (see, e.g.~\cite{Ghosh:2021axu}), implying a larger anisotropy for CDR.
This increase compensates for the Silk damping to some extent and is visible in the top right panel of figure~\ref{fig:TT}. Furthermore, the perturbation of CDR moves slower at sound speed $c_s = c/\sqrt{3} = 1/\sqrt{3}$ altering the gravitational drag on the photon-baryon perturbation. Thus, addition of CDR gives rise to a phase shift compared to FDR via Eq.~\eqref{eq:ps} and is seen in the top left panel of figure~\ref{fig:TT_frac}. Presence of interactions such as decoupling and recouping also non-trivially modify the phase-shift induce by DR (see for example Ref.~\cite{Montefalcone:2025ibh}).

The impact of decoupling and recoupling DR lies in between FDR and CDR, but adds additional scale dependence due to the different behavior of perturbations before and after the transition in the nature of DR. Decoupling DR behaves like CDR at early times (high $\ell$) and gradually approaches FDR after the transition, while recoupling DR has the opposite scale dependence (see the bottom two panels of figure~\ref{fig:TT} as well as the top left panel of figure~\ref{fig:TT_frac}).

\paragraph{Effects on Lensing.} Since the lensing of CMB photons is mostly caused by matter perturbations, the impact of DR on the CMB lensing power spectrum can be understood by its impact on the evolution of matter, in particular, CDM perturbations.

\begin{figure}
    \centering
    \includegraphics[width=0.45\linewidth]{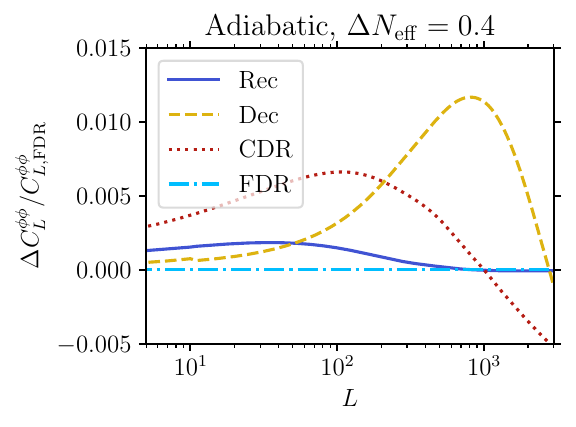}
    \includegraphics[width=0.45\linewidth]{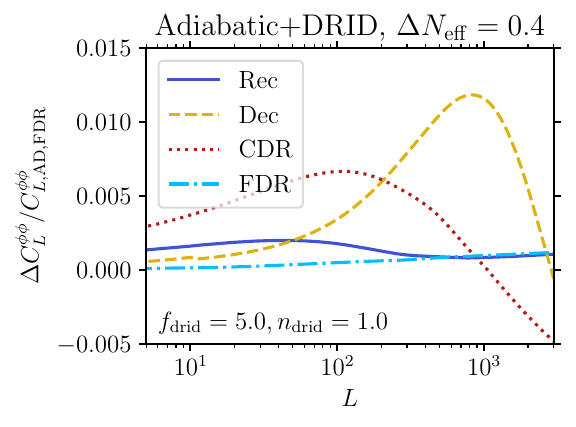}
    \includegraphics[width=0.45\linewidth]{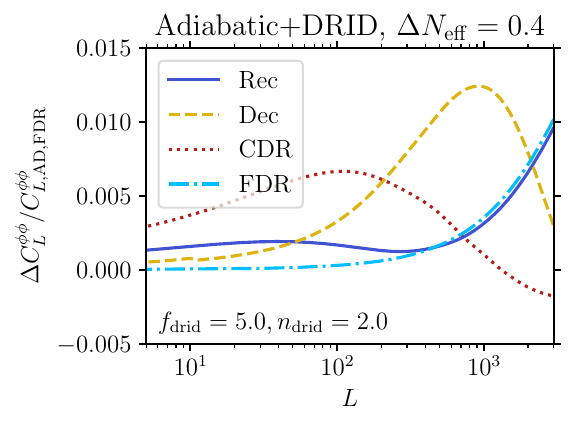}
    \includegraphics[width=0.45\linewidth]{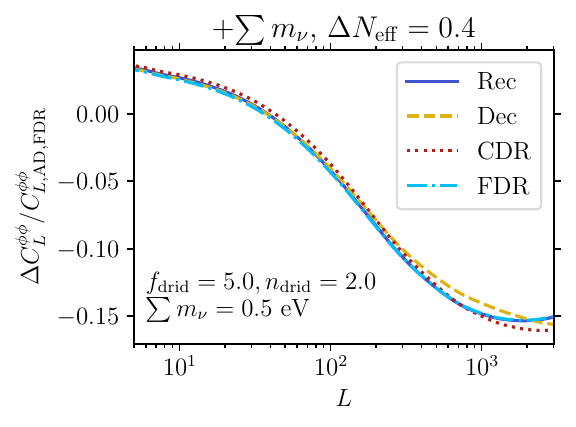}
    
    \caption{The fractional difference of $C^{\phi\phi}_L$ with respect to the adiabatic FDR case for different cases of DR scenarios with the same $\Delta N_{\rm eff}=0.4$. Here we show the effects of adiabatic and isocurvature IC, as well as massive neutrinos.}
    \label{fig:lensing_frac}
\end{figure}

Relative to the case of FDR, CDM modes re-entering the horizon during radiation domination experience a boost around horizon crossing in the presence of CDR.  They undergo subsequent suppression when they are deep inside the horizon due to rapid oscillations of interacting DR that suppress their time-averaged source term containing metric perturbations~\cite{Kreisch:2019yzn}. 
In more detail, from the gauge invariant perturbation in CDM $\zeta_c$, we observe CDM has a larger density fluctuation in the case of CDR, compared to FDR. Indeed, Eq.~\eqref{eq:adia_ic} shows, using $\zeta_c=-\eta + (1/3)\delta_c$, 
\es{}{
\zeta_{c,\rm FDR} = -1 + {5+4 R_\nu + 4 \rdr \over 12(15+4 R_\nu+4\rdr)}(k\tau)^2 - {1\over 12}(k\tau)^2,\\
\zeta_{c,\rm CDR} = -1 + {5+4 R_\nu \over 12(15+4 R_\nu)}(k\tau)^2 - {1\over 12}(k\tau)^2,
}
giving $|\zeta_{c,\rm CDR}| > |\zeta_{c,\rm FDR}|$ at the time of horizon re-entry.
Additionally, there is also sub-horizon dynamics just after mode re-entry that differ in CDR compared to FDR~\cite{Kreisch:2019yzn}.
Much after mode reentry, CDM modes experience a subsequent suppression when they are deep inside the horizon due to rapid oscillations of interacting DR that suppress the time-averaged source term.
The net effect is a suppression of high $k$ (or high $\ell$) modes relative to the case of FDR. When modes re-enter the horizon around matter-radiation equality, the suppression due to interacting DR becomes less in magnitude as the radiation density becomes subdominant. 
Therefore, the boost around horizon crossing leads to a net increase of the lensing spectrum for modes entering the horizon around equality.
For modes entering the horizon in the matter domination era (corresponds to small $\ell$), the impact of DR is negligible, and thus both FDR and interacting DR lead to the same results. Combining all the above effects, the CMB lensing spectrum with CDR relative to FDR exhibits a peak at intermediate scales and suppression at small scales~(see the top left panel of figure~\ref{fig:lensing_frac}). 

For decoupling DR, if the decoupling happens in the radiation domination era, the lensing spectrum has a peak higher than CDR at the scale corresponding to the decoupling time. This is because the modes around the peak experience a boost around horizon re-entry while DR is still interacting, but not much suppression later as DR becomes free-streaming inside the horizon.
Since DR soon becomes free-streaming after its decoupling, it approaches the case of FDR faster than CDR for small $\ell$. 
On the other hand, the recoupling case is the same as FDR for high $\ell$, but deviates (approaches the CDR case) at low $\ell$. The transition occurs around a scale set by the redshift of decoupling/recoupling given by Eq.~\eqref{eq:def_z_dec_rec} (see the top left panel of figure~\ref{fig:lensing_frac}). 
\begin{figure}
    \centering
    \includegraphics[width=0.45\linewidth]{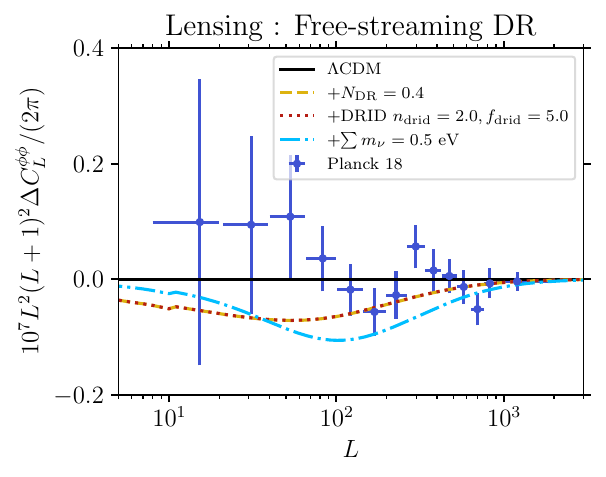}
    \includegraphics[width=0.45\linewidth]{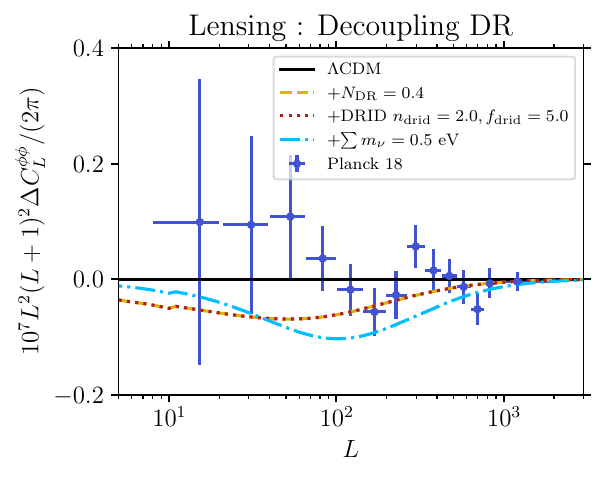}
    \includegraphics[width=0.45\linewidth]{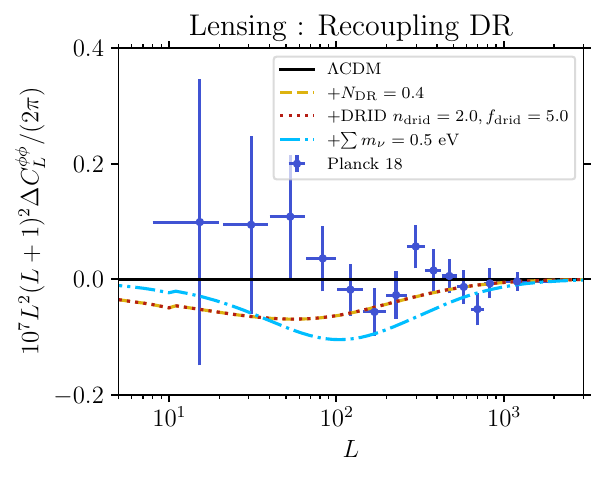}
    \includegraphics[width=0.45\linewidth]{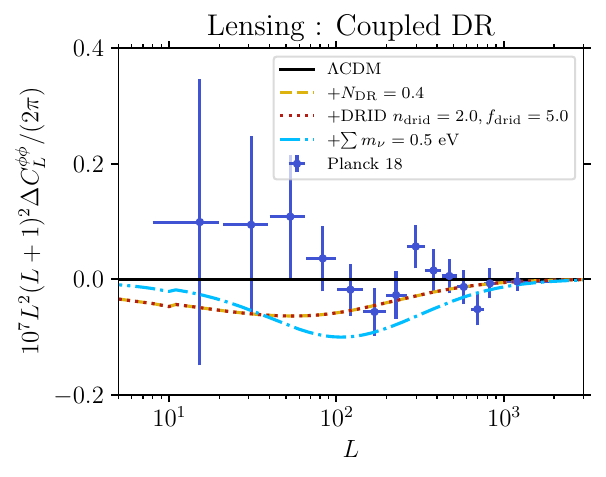}
    \caption{The CMB lensing spectrum for different cases of DR scenarios: FDR, CDR, decoupling and recoupling DR. For each type of DR, we show the effects of adiabatic and  isocurvature initial conditons, as well as massive neutrinos, comparing to the $\Lambda\rm CDM$ case. Data points are taken from Planck 2018~\cite{Planck:2018lbu}.}
    \label{fig:lensing}
\end{figure}

\paragraph{Isocurvature Initial Conditions.} We now move on to the effects of isocurvature. DR with isocurvature leads to an increase of CMB power spectra because of a positive contribution from isocurvature.\footnote{In this work, we assume isocurvature is \textit{uncorrelated} with curvature, and thus the isocurvature contribution to the power spectrum is generally positive.} 
For FDR, with a scale invariant primordial isocurvature power spectrum, $n_{\rm drid}=1$, $\Delta C^{\rm TT}_\ell/C^{\rm TT}_\ell$ decreases with $\ell$, relative to the adiabatic case, as seen in the top right panel of figure~\ref{fig:TT_frac}.
A primary reason is because the denominator $C_{\ell,{\rm AD, FDR}}^{\rm TT}$ increases with $\ell$ as we go from small $\ell$ to $\ell\sim 100$.
At the same time, as is well known, the locations of the acoustic peaks in a scenario with isocurvature IC are different compared to that with adiabatic IC~\cite{Hu:1996vq, Baumann:2015rya}.
This gives rise to the oscillations seen the top right panel of figure~\ref{fig:TT_frac}.
Other than these two features, the different DR species behave similar to the adiabatic scenario.
For the blue-tilted isocurvature power spectrum ($n_{\rm drid}=2$, bottom-left panels in figures~\ref{fig:TT_frac} and~\ref{fig:lensing_frac}), the enhancement in the resultant CMB spectra is more significant at high $\ell$ for all four kinds of DR, as expected. 

\paragraph{Neutrino Mass.} Massive neutrinos affect the CMB via multiple effects, including a late-ISW effect and changes to the distance to the surface of last scattering.
Since in figures~\ref{fig:TT} and \ref{fig:TT_frac}, $\theta_s$ is fixed, the effect of increasing $m_\nu$ manifests via a change in matter-dark energy equality, by driving $\Omega_m$ up.
This, in turn, leads to a decrease in the late-ISW effect and causes the dip at low-$\ell$ seen in figure~\ref{fig:TT_frac}.
As is well known, massive neutrinos leave a suppression in the CMB lensing spectrum and matter power spectrum at small scales due to their relativistic to non-relativistic transition.
This effect is manifest in figures~\ref{fig:lensing} and \ref{fig:lensing_frac}.
More detailed explanations can be found in, e.g., Refs.~\cite{Hu:1997mj,Dolgov:2002wy,Lesgourgues:2006nd,Hannestad:2010kz}
Since massive neutrinos and DR isocurvature have opposite effects on the lensing spectrum, varying the mass of neutrinos makes the constraint on $N_{\rm dr}$ weaker. We will show this clearly in section~\ref{subsec:AD_ISO}.

\section{Datasets and Methodology}\label{sec:datasets}
In this work, we used the following cosmological datasets to derive the constraint DR parameters. 

\begin{itemize}
    \item Planck: We used the baseline Planck temperature $(\TT: 2<\ell\lesssim 2500)$ and polarization $(\TE\EE: 2<\ell\lesssim 200 )$ likelihoods~\cite{Planck:2019nip}. We also include the Planck lensing likelihood~$(8< \ell \lesssim 400)$~\cite{Planck:2018lbu}.
    \item South Pole telescope (SPT): To include constraints from small-scale CMB observations, we include SPT-3G likelihood which includes temperature $(\TT: 750<\ell\lesssim 3000)$ and polarization $(\TE\EE: 300<\ell\lesssim 3000)$ measurements~\cite {SPT-3G:2021wgf,SPT-3G:2022hvq}.
    \item 
    Baryon Acoustic Oscillation (BAO): For the BAO measurements, we include the 6DF Galaxy survey~\cite{Beutler:2011hx}, SDSS-DR7 MGS data ~\cite{Ross:2014qpa}, and the BOSS measurement of BAO scale and $f\sigma_8$ from DR12 galaxy sample~\cite{BOSS:2016wmc}.
\item Pantheon+ (PAN): We use the Pantheon+ likelihood which contains $1701$ light curves
of $1550$ distinct Type Ia supernovae in the redshift range $0.001 < z < 2.26$~\cite{Brout:2022vxf}.
\item Pantheon+ \& SH0ES (PAN+SH0ES): Later in our analysis, we also include the measurements from SH0ES collaboration to assess the potential of DR models to solve the Hubble tension. We didn't use a direct likelihood of $H_0$, which has been shown to be incompatible with the use of SN light curves likelihood like Pantheon+~\cite{Benevento:2020fev,Camarena:2021jlr,Efstathiou:2021ocp}. Rather, we used the Pantheon + SH0ES likelihood, which uses SH0ES cepheid host distance calibration for Pantheon+ supernovae samples~\cite{Brout:2022vxf,Riess:2021jrx}.
\end{itemize}
\section{MCMC Results}
We used \texttt{Montepython}~\cite{Brinckmann:2018cvx,Audren:2012wb} to perform the MCMC runs and used \texttt{GetDist}~\cite{Lewis:2019xzd} for the analysis of the MCMC chains. We used the Metropolis-Hastings algorithm~\cite{Metropolis:1953am,Hastings:1970aa} for the MCMC runs. Several parameter sets in our analysis suffer from large degeneracies, for example, $\Ndr$ and $N_\nu$ in the FDR case, whose effects on background cosmology are very similar. Due to this reason, we need to run chains for a significantly longer time for desired convergence. In addition, some parameters become unconstrained in certain limits, such as $m_\nu$ in the limit $N_\nu \to 0$.
These effects also lead to slower convergence.  We adopted the Gelman-Rubin convergence criterion of $R-1 < 0.1$~\cite{Gelman:1992zz} for the MCMC runs. 

In the following part, we will discuss the results of the MCMC analysis. Before we go into model-specific details, we lay out the definition of several quantities used in figures and tables. In each table, we show the improvement of the $\chi^2$ over the baseline $\Lambda$CDM model with massless neutrinos and fixed $\Neff = 3.044$~\cite{Akita:2020szl, Froustey:2020mcq, Bennett:2020zkv}. We also mention the Akaike Information Criterion (AIC) of model comparison, defined as
\begin{equation}
    \label{eq:AIC}
    {\rm AIC} = \Delta \chi^2 + 2 (N_\mathcal{M} - N_{\Lambda{\rm CDM}})\;,
\end{equation}
where $N_\mathcal{M}$ is the number of free parameters in model $\mathcal{M}$. Note that the minimum number of extra parameters for models in this analysis is three ($N_{\rm dr}, N_\nu, m_\nu$). Therefore, $N_\mathcal{M}$ can be significantly larger than $N_{\Lambda{\rm CDM}} = 6$. We didn't fix any of the new parameters to any special value; rather, we varied all those along with the $\Lambda$CDM parameters to scan the whole parameter space. Although some cases we consider will have a positive AIC, our goal in the present paper is to understand the interplay and all possible degeneracies among the various parameters.

Due to strong parameter degeneracies, the $\chi^2$ computed from the best fit MCMC sample is prone to being less accurate. Therefore, to find the minimum $\chi^2$ for each model, we performed a dedicated run using \texttt{Montepython}'s `minimization' function with the MCMC best fit as the starting point. \texttt{Montepython}'s `minimization' algorithm uses the python package \texttt{scipy}~\cite{Virtanen:2019joe}. The difference in the value of $\chi^2$ can be substantial $(\sim 20)$ between this method and the MCMC best fit.\footnote{We found out that the `Powell' method of minimization in \texttt{scipy} overall works the best for our case. However, we note that the \texttt{Montepython}'s (\texttt{scipy})`minimization' is not suitable for large-dimensional parameter space. In a handful of cases, the `minimization' fails to provide a best fit point. In those scenarios, we rerun the `minimization' 5 times, subsequently reducing the tolerance (`tol') from $10^{-8}$ to $10^{-4}$ till we get a best fit. If all 5 runs fail, we use the smallest $\chi^2$ point traversed during the minimizing runs before failing as the minimum $\chi^2$. Thus, in those cases, the minimum $\chi^2$ quotes can be larger than the true global minimum. Although this method may fail to find the true global minima, the minimum $\chi^2$ in this method is still smaller than the naive MCMC best fit $\chi^2$ for all the cases studied in this paper. Since the goal of this work is to derive parameter constraint, we stick to this method for quoting the value of $\chi^2$ for the tables. Profile likelihood codes based on a frequentist approach, such as \texttt{Procoli}~\cite{Karwal:2024qpt}, \texttt{PROSPECT}~\cite{Holm:2023uwa} are better suited to find the global minima, however, their use is beyond the scope of the paper.}
\label{sec:MCMC_results}
\subsection{ Adiabatic Initial Conditions}\label{subsec:AD}
\begin{figure}
    \centering
    \includegraphics[width=0.49\linewidth]{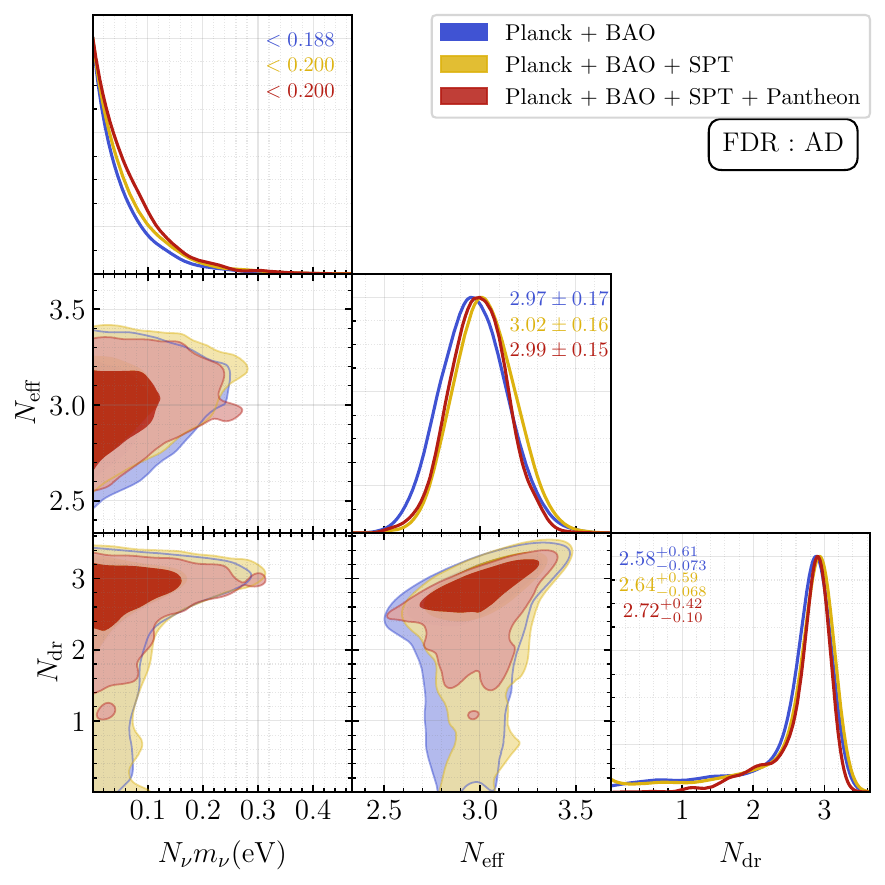}
    \includegraphics[width=0.49\linewidth]{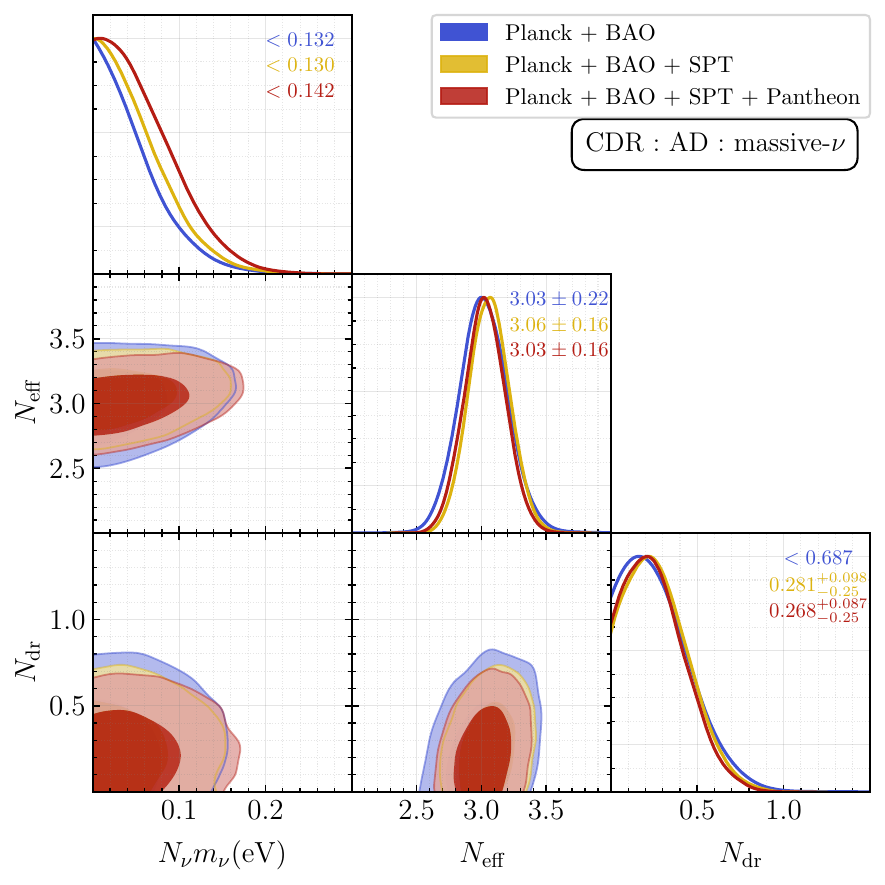}
    \caption{Triangle plots for parameters for FDR (left) and CDR (right) with adiabatic perturbations. Constraints are noted at the upper corners of the 1D contours. The error bars are at $1\sigma$ and the upper limits are mentioned at $95\%$ confidence level (C.L.). }
    \label{fig:adia-fdr-cdr}
\end{figure}
First, we derive the cosmological constraints for different DR interaction scenarios carrying adiabatic perturbations only. In common to all the scenarios studied in this work, we added at minimum \emph{three} more parameters on top of the base $\Lambda$CDM model: the number of effective neutrino species $N_\nu$, neutrino mass $m_\nu$, and effective number of DR species $N_{\rm dr}$. In this work, we choose the neutrinos to have a degenerate mass.\footnote{We implemented massive neutrinos via the `ncdm' module in \texttt{CLASS}. We set the \texttt{CLASS} parameter \texttt{N\_{ncdm}} = 1 and varied \texttt{deg\_ncdm} to control the neutrino number density. For massless DR, we use the `idr' module and made necessary changes to incorporate DR self interactions.} Thus, the total $\Neff = N_\nu + N_{\rm dr}$ contained contributions from both these species. We used the following flat priors for these quantities: $N_{\rm dr} \in [0,\infty)$\footnote{Prior upper boundary set to $\infty$ means that it was not specified, so, there was no hard prior upper boundary.}, $N_\nu \in [0,\infty)$ and $m_\nu \in [0, 5~{\rm eV}]$.

To begin with, we will focus on two simpler scenarios where the DR is free-streaming (FDR) as neutrinos in $\Lambda$CDM, or fluid-like coupled DR (CDR) where strong interaction hinders DR free-streaming till today. Thus, in these two cases, the additional number of parameters in addition to the $\Lambda$CDM model is three.   Figure~\ref{fig:adia-fdr-cdr} shows the marginalized constraints on the number densities of DR and neutrinos, as well as the neutrino mass. We plot a quantity $N_\nu m_\nu  \equiv \omega_{\nu,0} \times  93.14 \ {\rm eV}$ with $\omega_{\nu,0} \equiv \bar{\rho}_{\nu,0} h^2/(3 H_0^2\mpl^2)$ and $h\equiv H_0/(100~{\rm km/s}/{\rm Mpc})$.
Thus, $N_\nu m_\nu$ quantifies the effective total neutrino mass\footnote{This is equivalent to the variable \texttt{m\_ncdm\_tot} in \texttt{CLASS}.} and it is proportional to the physical non-relativistic energy density of neutrinos, which is the physical quantity directly constrained by the data.  Figure~\ref{fig:adia-fdr-cdr} (left) shows the constraint on the FDR case. In this case, both the neutrino and DR are free-streaming, but the neutrino has a non-zero mass whereas DR is massless. The plot shows that the Planck data prefer massless DR over massive neutrinos with the total $N_{\rm eff}$ being close to $3$. This is because the Planck CMB data exhibit an enhanced large-scale lensing power, which drives the preference for zero (rather, negative) mass for neutrinos in $\Lambda$CDM~\cite{Craig:2024tky,Green:2024xbb,Loverde:2024nfi,Lynch:2025ine,Graham:2025fdt}. In addition, the BAO data favor lower matter densities, which also drive the preference for smaller neutrino masses~\cite{Loverde:2024nfi,Lynch:2025ine}. Due to this generic preference for massless non-photon species in CMB and BAO data, massless DR is preferred over massive neutrinos.   

\begin{figure}
    \centering
    \includegraphics[width=0.49\linewidth]{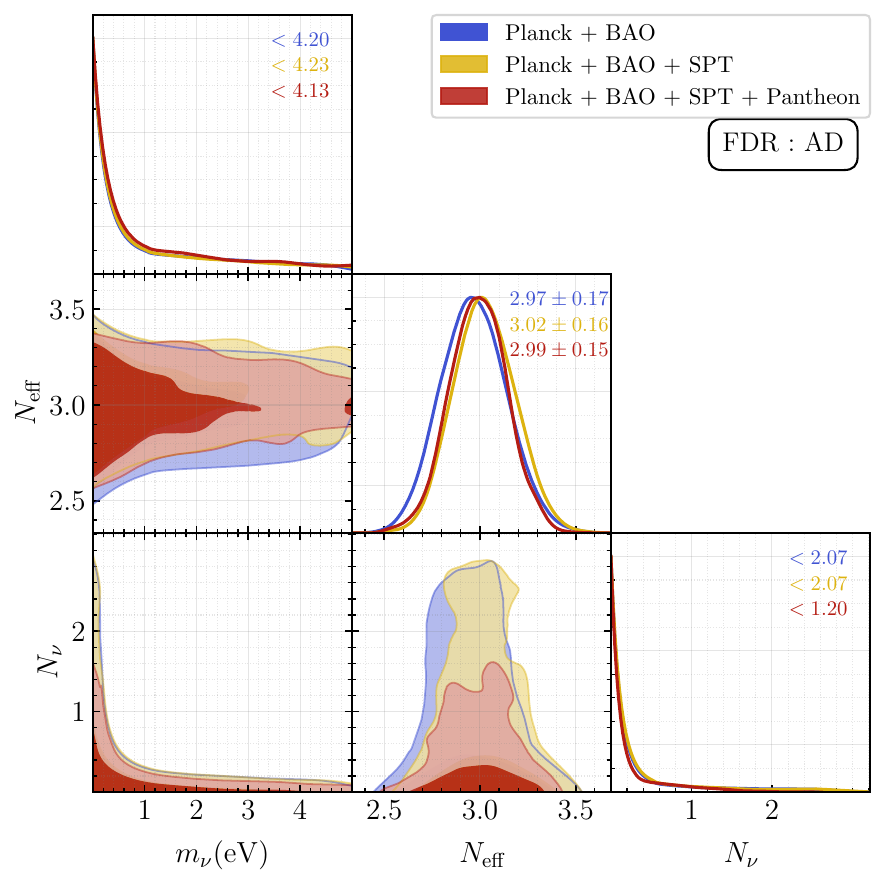}
    \includegraphics[width=0.49\linewidth]{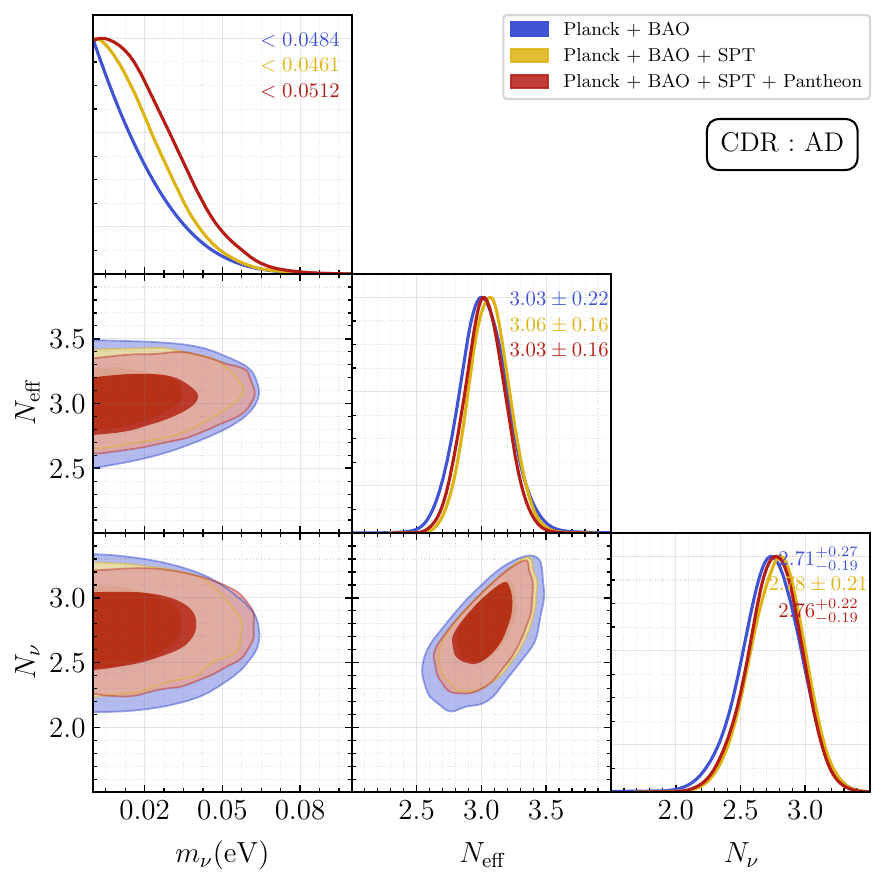}
    \caption{Triangle plots for neutrino mass and abundance for FDR (left) and CDR (right) with adiabatic perturbations. Constraints are noted at the upper corners of the 1D contours. The error bars are at $1\sigma$ and the upper limits are mentioned at $95\%$ confidence level (C.L.). }
    \label{fig:N_prop}
\end{figure}

The addition of the SPT and Pantheon+ dataset does not affect the preference of massless DR over massive neutrinos, but slightly relaxes the $N_\nu m_\nu$ bound. Specifically for the Pantheon+ dataset, the relaxation of $N_\nu m_\nu $ is related to the preference for larger matter fraction $(\Omega_m)$, which supports slightly positive neutrino mass~\cite{Brout:2022vxf,Loverde:2024nfi}. A small $N_{\nu}$ necessarily means that constraints on neutrino mass are relaxed in this scenario. Figure~\ref{fig:N_prop} (left) shows the constraint on $N_\nu$ and the individual neutrino mass $ m_\nu$. Due to the very small $N_\nu$, the  $ m_\nu$ upper bound is drastically relaxed to the upper prior boundary $\sim 5\, {\rm eV}$ at 95\% C.L.  Note that terrestrial experiments already set a much stronger upper limit of $m_\nu < 0.45$ eV (90\% C.L.)~\cite{KATRIN:2024cdt}. Thus, our analysis shows that when $N_\nu$ is allowed to vary and if additional massless species are present, the neutrino mass bound from cosmology can be significantly relaxed (see for example~\cite{PhysRevLett.93.121302,Serpico:2007pt,Chacko:2019nej,Chacko:2020hmh,Escudero:2020ped,FrancoAbellan:2021hdb}).

For the fluid-like CDR case, the DR has properties different from those of neutrinos. Compared to the FDR case, the CMB spectrum with CDR has less damping at high multipoles and a different acoustic phase shift, which can be seen from figure~\ref{fig:TT_frac}. The cosmological data prefer most of the additional radiation in the Universe to be free-streaming at recombination. Thus, massive neutrinos make up most of $N_{\rm eff}$ in this case, unlike the FDR one. In this case, the total $N_{\rm eff}$ is slightly larger than in the FDR case, which can be attributed to the effects of CDR on the damping tail~\cite{Blinov:2020hmc,Ghosh:2021axu}. The bound on $N_\nu m_\nu$ is also tighter in the CDR scenario. Since $N_\nu$ can account for almost all $N_{\rm eff}$ in this case, the bound on $m_\nu$ is much tighter as can be seen from figure~\ref{fig:N_prop}. The constraints on the parameters in both FDR and CDR cases are shown in table~\ref{tab:fdr-cdr}.  
\renewcommand{\arraystretch}{1.3}
\begin{table}[t]
    \centering
\resizebox{\textwidth}{!}{
    \begin{tabular}{|c||c|c||c|c||c|c||}
    \hline
         &\multicolumn{2}{c||}{Planck + BAO} & \multicolumn{2}{c||}{Planck + BAO + SPT}  & \multicolumn{2}{c||}{Planck + BAO + SPT + PAN} \\
         \hline
         & FDR & CDR & FDR & CDR & FDR & CDR \\
        \hline
$ 10^2 \omega_{b}$ & $2.233\pm 0.019$ & $2.244^{+0.018}_{-0.023}$ & $2.234\pm 0.018$ & $2.242\pm 0.018$ & $2.229\pm 0.017$ & $2.238\pm 0.018$\\
\hline
$ \omega{}_{\rm cdm }$ & $0.1182\pm 0.0028$ & $0.1197^{+0.0029}_{-0.0033}$ & $0.1190\pm 0.0026$ & $0.1203^{+0.0025}_{-0.0028}$ & $0.1188\pm 0.0025$ & $0.1200\pm 0.0027$\\
\hline
$ 100\theta{}_{s }$ & $1.04210\pm 0.00049$ & $1.04297^{+0.00057}_{-0.00087}$ & $1.04197\pm 0.00044$ & $1.04273^{+0.00059}_{-0.00073}$ & $1.04200^{+0.00040}_{-0.00047}$ & $1.04276^{+0.00057}_{-0.00072}$\\
\hline
$ 10^{10}{\cal P}_{\mathcal{RR}}^{(1)}$ & $23.66\pm 0.47$ & $23.74\pm 0.55$ & $23.52\pm 0.47$ & $23.59\pm 0.46$ & $23.61\pm 0.43$ & $23.68\pm 0.46$\\
\hline
$ 10^{10}{\cal P}_{\mathcal{RR}}^{(2)}$ & $20.42\pm 0.39$ & $20.10\pm 0.64$ & $20.44^{+0.35}_{-0.39}$ & $20.11\pm 0.42$ & $20.34^{+0.33}_{-0.38}$ & $20.07\pm 0.42$\\
\hline
$ N_{\rm dr}$ & $2.58^{+0.61}_{-0.073}$ & $< 0.362$ & $2.64^{+0.59}_{-0.068}$ & $0.281^{+0.098}_{-0.25}$ & $2.72^{+0.42}_{-0.10}$ & $0.268^{+0.087}_{-0.25}$\\
\hline
$ N_\nu$ & $< 0.219$ & $2.71^{+0.27}_{-0.19}$ & $< 0.235$ & $2.78\pm 0.21$ & $< 0.203$ & $2.76^{+0.22}_{-0.19}$\\
\hline
$ m_\nu ({\rm eV})$ & $< 1.52$ & $< 0.0214$ & $< 1.44$ & $< 0.0230$ & $< 1.52$ & $< 0.0268$\\
\hline
\hline
$ N_{\rm eff}$ & $2.97\pm 0.17$ & $3.03\pm 0.22$ & $3.02\pm 0.16$ & $3.06\pm 0.16$ & $2.99\pm 0.15$ & $3.03\pm 0.16$\\
\hline
$ N_\nu m_\nu (\rm eV)$ & $< 0.0679$ & $< 0.0581$ & $< 0.0761$ & $< 0.0637$ & $< 0.0807$ & $< 0.0741$\\
\hline
$ H_0 ({\rm km/s/Mpc})$ & $67.2\pm 1.1$ & $67.8\pm 1.5$ & $67.5\pm 1.1$ & $67.9\pm 1.1$ & $67.15\pm 0.97$ & $67.5\pm 1.0$\\
\hline
$ \sigma_8$ & $0.808^{+0.013}_{-0.011}$ & $0.806^{+0.013}_{-0.0096}$ & $0.808^{+0.013}_{-0.0099}$ & $0.807^{+0.011}_{-0.010}$ & $0.807^{+0.013}_{-0.0098}$ & $0.805^{+0.012}_{-0.010}$\\
\hline
$ 10^{-9}A_{s }$ & $2.096\pm 0.034$ & $2.070\pm 0.057$ & $2.095\pm 0.033$ & $2.069\pm 0.037$ & $2.088\pm 0.033$ & $2.067\pm 0.037$\\
\hline
$ n_{s }$ & $0.9623\pm 0.0069$ & $0.9574\pm 0.0090$ & $0.9641\pm 0.0068$ & $0.9592\pm 0.0073$ & $0.9619\pm 0.0063$ & $0.9577\pm 0.0072$\\
\hline
$\Delta \chi^2$ & $-0.9$ & $-0.3$ & $1.3$ & $-0.8$ & $-0.5$ & $-1.1$\\
\hline
AIC & $5.1$ & $5.7$ & $7.3$ & $5.2$ & $5.5$ & $4.9$\\
\hline
    \end{tabular}
}
    \caption{Marginalized constraints in the primary and derived cosmological parameters (below double line) for the FDR and CDR case for all datasets. The error bars are at $1\sigma$ and upper limits are at $68\%$ C.L. }
    \label{tab:fdr-cdr}
\end{table}

\begin{figure}[t]
    \centering
    \includegraphics[width=0.49\linewidth]{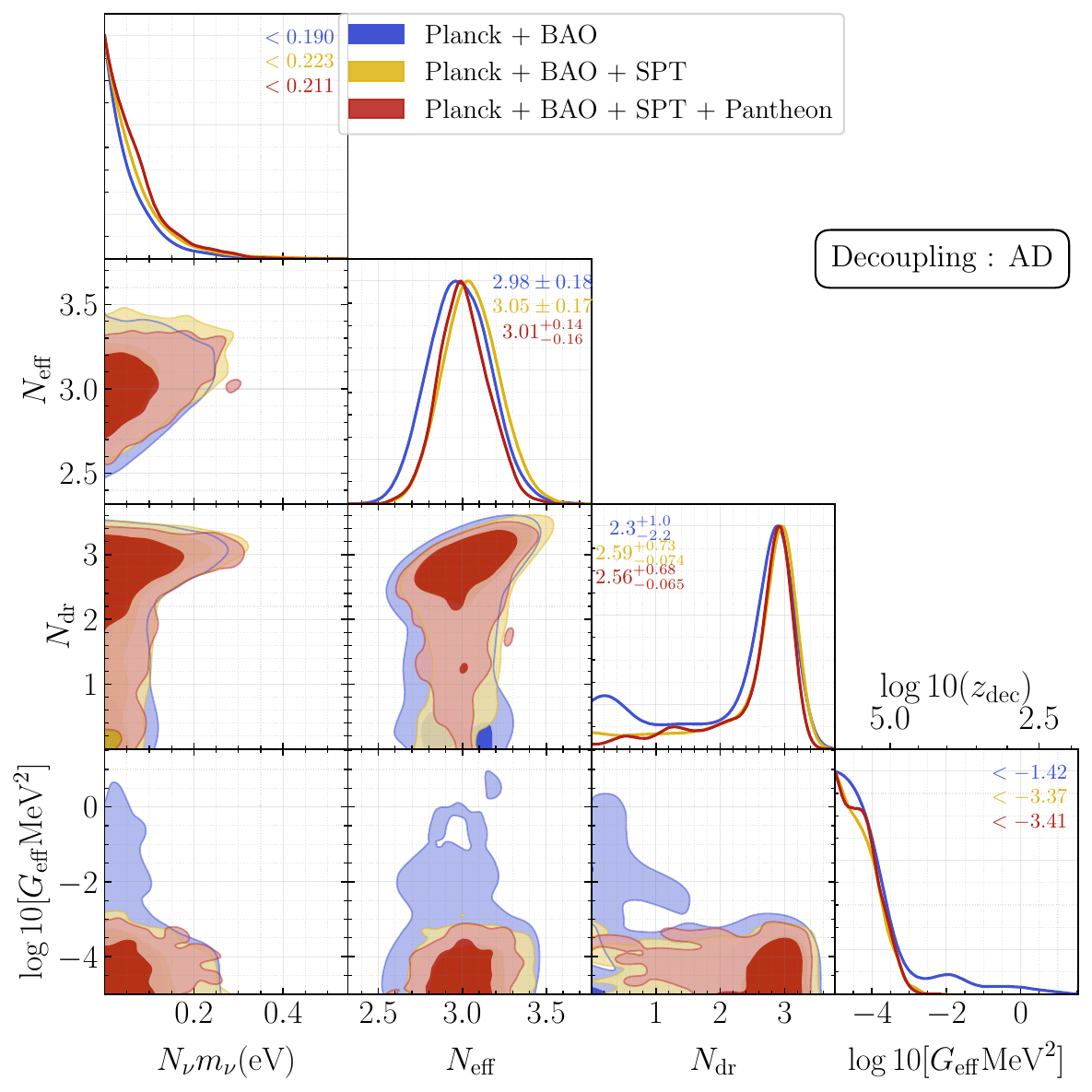}
    \includegraphics[width=0.49\linewidth]{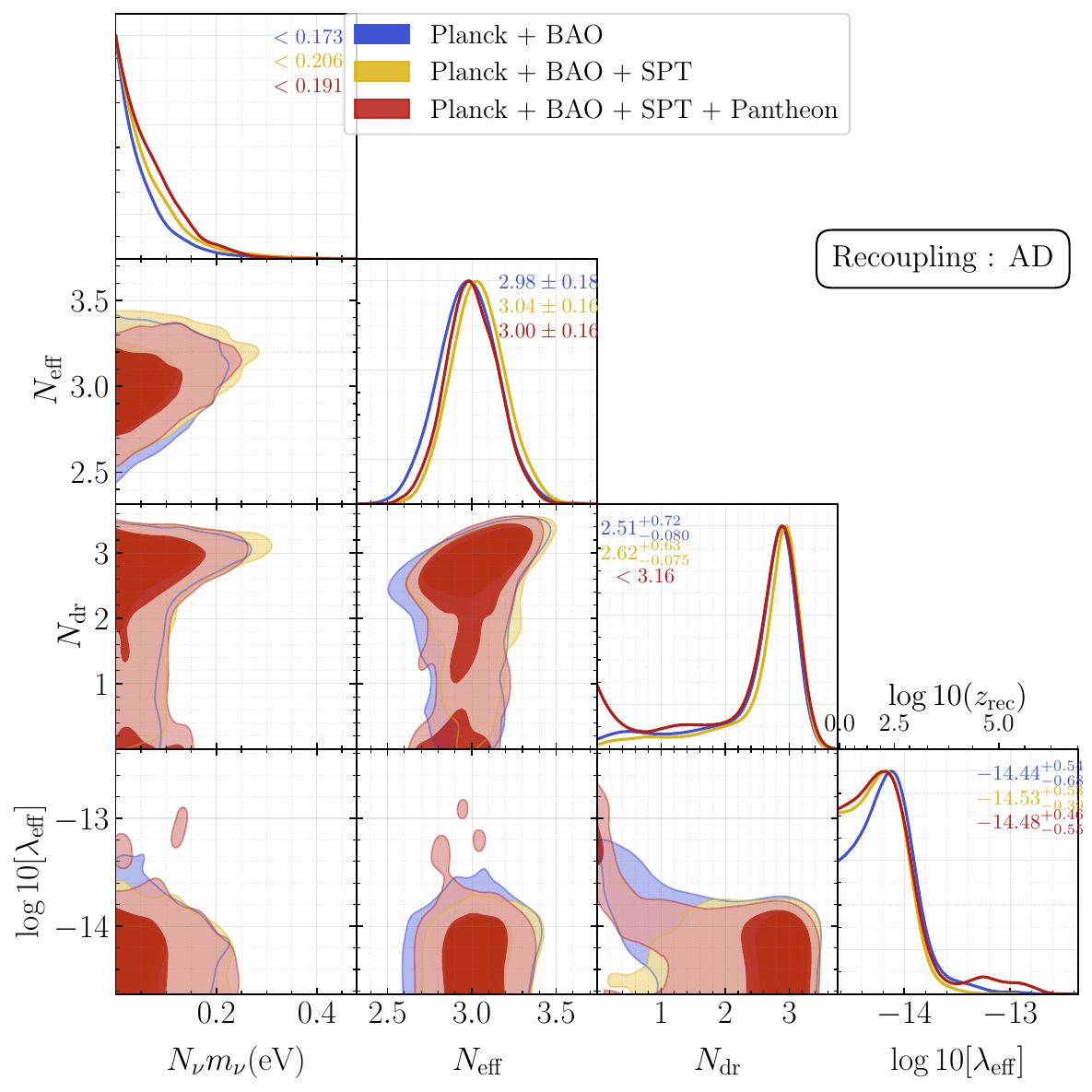}
    \caption{Triangle plots for parameters with decoupling (left) and recoupling (right) DR carrying adiabatic perturbations only. Constraints are noted at the upper corners of the 1D contours. The error bar are at $1\sigma$ and the upper limits are mentioned at $95$\% C.L..}
    \label{fig:adia-decoup-recoup}
\end{figure}

 Next, we turn to cases where DR can transition between FDR and CDR due to the presence of interactions. 
 Cosmological datasets are sensitive to these transitions since the DR free-streaming properties are different across the transition redshift and induce different characteristic features on the CMB spectrum, as shown in the previous sections. We study both the cases of decoupling and recoupling DR and constrain the coupling parameters. 
 While performing the MCMC, we impose a flat prior on $\lgeff \in [-5.0, 2.0]$ for decoupling DR and  $\lleff \in [-15.3, -11.8]$\footnote{In \texttt{drisoCLASS} code the recoupling interaction is coded in a unified way with the same interaction variable $\log_{10}(G_{\rm eff})$ as in the decoupling case with proper temperature scaling of the interaction strength. The relation between these two variables for the recouping case are: $\lleff = (\log_{10}(G_{\rm eff}) - 25.56)/2$. The above specific prior range is a result of setting this following prior on $\log_{10}(G_{\rm eff}) \in [-5.0, 2.0]$ for the recoupling case.} for recoupling DR. Figure~\ref{fig:adia-decoup-recoup} shows the constraint on decoupling DR on the left and recoupling DR on the right.
 
 In the case of decoupling, the DR is coupled at an earlier time, with decoupling happening at $\Gamma \sim H$ and after that DR is free-streaming. The opposite happens in the case of recoupling, where $\Gamma \ll H$ at an earlier time and thus DR is free-streaming. At late times, when $\Gamma \gtrsim H$, the DR transitions to a fluid-like species. The coupling strengths in these two cases are given by $G_{\rm eff}$ and $\lambda_{\rm eff}$, which are defined in Eq.~\eqref{eq:ave_rate_decoupling} and \eqref{eq:ave_rate_recoupling}, respectively. We quantify the redshift of decoupling and recoupling as $ \Gamma_\textrm{eff}(\Geff,z_{\rm dec})=H(z_{\rm dec})$ and $ \Gamma_\textrm{eff}(\lambda_{\rm eff},z_{\rm rec})=H(z_{\rm dec})$, respectively (see Eq.~\eqref{eq:def_z_dec_rec}). 
 
 The 1D marginalized posterior of $\lgeff$ in figure~\ref{fig:adia-decoup-recoup} (left panel) for decoupling DR shows that data prefer a small value of $\Geff$, which supports early decoupling. The data prefer DR to be free-streaming rather than coupled at scales relevant for CMB. This is expected from the previous CDR analysis where the DR fraction of the total $\Neff$ was quite smaller compared to a similar FDR analysis. Since the DR is mostly free-streaming, the constraints for other parameters are closer to the previous FDR analysis and massless DR contributes most to the $N_{\rm eff}$ compared to the massive neutrinos. The addition of SPT and the Pantheon data makes the bound of $\lgeff$ stronger. The top axis of the 1D posterior of $\lgeff$ shows the corresponding values of $z_{\rm dec}$. The constraint of $\lgeff$ translates to $z_{\rm dec} \gtrsim 10^5$ at 95 \% C.L..

 Figure~\ref{fig:adia-decoup-recoup} (right panel) shows the case for recoupling DR. In this case, data also prefer a small value of $\lleff$, which supports that DR is mostly free-streaming at CMB length scales. The constraint on $\lleff$ can be translated to $z_{\rm rec} \lesssim 10^3 $ at 95\% C.L.. Similar to the decoupling case, the massless DR contributes most to the total $\Neff$ and the bound on $N_\nu m_\nu$ is slightly tighter than the decoupling case. In summary, the cosmological data prefer the bulk of additional radiation species to be mostly free-streaming. Thus, even in the presence of coupling, the marginalized posterior peaks at the small coupling limit. However, the constraints on the coupling get weaker for smaller $\Ndr$ which can be seen from the 2D plots in figures~\ref{fig:adia-decoup-recoup}.
 The constraints on the parameters in both cases are shown in table~\ref{tab:decoup-recoup}.

\begin{table}[]
    \centering
\resizebox{\textwidth}{!}{
    \begin{tabular}{|c||c|c||c|c||c|c||}
    \hline
         &\multicolumn{2}{c||}{Planck + BAO} & \multicolumn{2}{c||}{Planck + BAO + SPT}  & \multicolumn{2}{c||}{Planck + BAO + SPT + PAN} \\
         \hline
         & Decoupling & Recoupling & Decoupling & Recoupling & Decoupling & Recoupling \\
        \hline
$ 10^2 \omega_{b}$ & $2.233\pm 0.019$ & $2.234\pm 0.019$ & $2.234^{+0.017}_{-0.019}$ & $2.235\pm 0.017$ & $2.230\pm 0.016$ & $2.230\pm 0.017$\\
\hline
$ \omega{}_{\rm cdm }$ & $0.1185\pm 0.0030$ & $0.1184\pm 0.0029$ & $0.1195\pm 0.0027$ & $0.1193^{+0.0024}_{-0.0027}$ & $0.1190\pm 0.0024$ & $0.1191\pm 0.0025$\\
\hline
$ 100\theta{}_{s }$ & $1.04220^{+0.00046}_{-0.00064}$ & $1.04223\pm 0.00051$ & $1.04192\pm 0.00044$ & $1.04204\pm 0.00045$ & $1.04199\pm 0.00043$ & $1.04209\pm 0.00045$\\
\hline
$ 10^{10}{\cal P}_{\mathcal{RR}}^{(1)}$ & $23.76\pm 0.49$ & $23.50\pm 0.52$ & $23.58\pm 0.48$ & $23.36\pm 0.51$ & $23.70\pm 0.46$ & $23.49^{+0.52}_{-0.45}$\\
\hline
$ 10^{10}{\cal P}_{\mathcal{RR}}^{(2)}$ & $20.34\pm 0.43$ & $20.43\pm 0.39$ & $20.42\pm 0.37$ & $20.45^{+0.34}_{-0.40}$ & $20.34^{+0.34}_{-0.40}$ & $20.36\pm 0.39$\\
\hline
$ N_{\rm dr}$ & $2.3^{+1.0}_{-2.2}$ & $2.51^{+0.72}_{-0.080}$ & $2.59^{+0.73}_{-0.074}$ & $2.62^{+0.63}_{-0.075}$ & $2.56^{+0.68}_{-0.065}$ & $< 2.93$\\
\hline
$ N_\nu$ & $< 0.573$ & $< 0.312$ & $< 0.315$ & $< 0.276$ & $< 0.307$ & $< 0.483$\\
\hline
$ \log10[G_{\rm eff} ~{\rm MeV}^2]$ & $< -3.86$ & $--$ & $< -4.10$ & $--$ & $< -4.08$ & $--$\\
\hline
$ \log10[\lambda_{\rm eff}]$ & $--$ & $-14.44^{+0.54}_{-0.68}$ & $--$ & $-14.53^{+0.53}_{-0.39}$ & $--$ & $-14.48^{+0.46}_{-0.55}$\\
\hline
$ m_\nu ({\rm eV})$ & $< 1.26$ & $< 1.26$ & $< 1.44$ & $< 1.37$ & $< 1.81$ & $< 1.10$\\
\hline
\hline
$ N_{\rm eff}$ & $2.98\pm 0.18$ & $2.98\pm 0.18$ & $3.05\pm 0.17$ & $3.04\pm 0.16$ & $3.01^{+0.14}_{-0.16}$ & $3.00\pm 0.16$\\
\hline
$ N_\nu m_\nu (\rm eV)$ & $< 0.0692$ & $< 0.0640$ & $< 0.0796$ & $< 0.0788$ & $< 0.0814$ & $< 0.0844$\\
\hline
$ H_0 ({\rm km/s/Mpc})$ & $67.3\pm 1.2$ & $67.3\pm 1.1$ & $67.7\pm 1.1$ & $67.6\pm 1.0$ & $67.26^{+0.87}_{-1.0}$ & $67.20^{+0.94}_{-1.1}$\\
\hline
$ \sigma_8$ & $0.809^{+0.013}_{-0.011}$ & $0.809^{+0.013}_{-0.011}$ & $0.809^{+0.014}_{-0.0099}$ & $0.809^{+0.014}_{-0.0099}$ & $0.808^{+0.013}_{-0.010}$ & $0.807^{+0.013}_{-0.011}$\\
\hline
$ 10^{-9}A_{s }$ & $2.090\pm 0.037$ & $2.094\pm 0.034$ & $2.094\pm 0.032$ & $2.093^{+0.030}_{-0.034}$ & $2.089^{+0.030}_{-0.036}$ & $2.088\pm 0.034$\\
\hline
$ n_{s }$ & $0.9603^{+0.0082}_{-0.0070}$ & $0.9642\pm 0.0076$ & $0.9632\pm 0.0070$ & $0.9659\pm 0.0071$ & $0.9608\pm 0.0065$ & $0.9635\pm 0.0068$\\
\hline
$ \log10 (z_{\rm dec})$ & $5.26^{+0.68}_{-0.15}$ & $--$ & $5.52^{+0.42}_{-0.21}$ & $--$ & $5.51^{+0.41}_{-0.21}$ & $--$\\
\hline
$ \log10 (z_{\rm rec})$ & $--$ & $-0.1^{+4.2}_{-6.3}$ & $--$ & $-0.6^{+4.3}_{-2.0}$ & $--$ & $-0.5^{+4.5}_{-6.9}$\\
\hline
$\Delta \chi^2$ & $-2.0$ & $-2.2$ & $-0.4$ & $-0.9$ & $-0.7$ & $1.0$\\
\hline
AIC & $6.0$ & $5.8$ & $7.6$ & $7.1$ & $7.3$ & $9.0$\\
\hline
    \end{tabular}
}
    \caption{Marginalized constraints in the primary and derived cosmological parameters (below double line) for the decoupling and recoupling DR for all datasets. The error bars are at $1\sigma$ and upper limits are at $68\%$ C.L.. }
    \label{tab:decoup-recoup}
\end{table}

\subsection{ Adiabatic + Isocurvature Initial Conditions}
\label{subsec:AD_ISO}
We perform a similar analysis to the previous section in the presence of DR isocurvature (DRID). Figure~\ref{fig:TT} showed the effects of DRID on the CMB spectrum. 
As can be seen from table~\ref{tab:iso_ic}, the following density and metric perturbations of DRID in the synchronous gauge have a universal feature:
\begin{equation}
    \label{eq:iso_scaling}
    \delta_\gamma, \theta_\gamma, \delta_b, \delta_c, \delta_\nu,\theta_\nu, h, \eta \propto \dfrac{R_{\rm dr}}{1 - R_{\rm dr}} \approx R_{\rm dr} \propto N_{\rm dr}~~~~(\textrm{for}~ R_{\rm dr}\ll1).
\end{equation}
In the last equation we have used $R_{\rm dr}$ is small compared to $R_\nu$ and $R_\gamma \equiv 1 - R_{\rm dr} - R_\nu$. This is an excellent approximation for CDR where $N_{\rm dr}$ is expected to be small compared to the total $\Neff$. Note that, for FDR $R_{\rm dr}$ can be as large as $R_{\rm dr} \lesssim 0.4$ if DR contributes to the majority of the total $\Neff$. Therefore Eq.\eqref{eq:iso_scaling} holds approximately for FDR up to a factor of order unity. Following the scaling of Eq.~\eqref{eq:iso_scaling} and our definition of the DRID power spectrum in Eq.~\eqref{eq:P_iso}, the physical combination $N_{\rm dr}^2 A_{\rm drid} = N_{\rm dr}^2 f_{\rm drid}^2 A_s$ determines the magnitude of the effects of DRID on CMB observables. Since $N_{\rm dr}$ can take a wide range of values (close to zero for CDR and close to $3$ for FDR), we decided to vary $N_{\rm dr}^2 f_{\rm drid}^2$ as a primary parameter, following Ref.~\cite{Ghosh:2021axu}. For the MCMC runs, we employed the `two scale' parametrization where a single power law spectrum with amplitude $A_s(A_{\rm drid})$ defined at the pivot scale $k_\ast$ and tilt $n_s(n_{\rm drid})$ can be defined by two power spectrum amplitudes $\mathcal{P}_{\mathcal{RR}}^{(1,2)} (\mathcal{P}_{\mathcal{II}}^{(1,2)})$  defined at two scales $k_1,k_2$. For adiabatic perturbations, we varied $P_{\mathcal{RR}}^{(1)}$ and $P_{\mathcal{RR}}^{(2)}$ as primary parameters. $A_s$ and $n_s$ are derived in the following way
\begin{equation}
    \label{eq:nsAs_def}
    n_s = 1 + \dfrac{\ln {\cal P}_{\mathcal{RR}}^{(1)} - \ln {\cal P}_{\mathcal{RR}}^{(2)}}{\ln k_1 - \ln k_2}, \qquad A_s = {\cal P}_{\mathcal{RR}}^{(1)} \exp\left[(n_s -1 )\ln\left(k_\ast \over k_1\right)\right]\;.
\end{equation}
Following the above discussion of the DRID case, it is more convenient to vary $N_{\rm dr}^2{\cal P}_{\mathcal{II}}^{(1)}$ and $N_{\rm dr}^2 {\cal P}_{\mathcal{II}}^{(2)}$ and thus we treat them as primary parameters with a flat prior $[0, {\rm \infty})$. We report the tilt $n_{\rm drid}$ and the scaled amplitude fraction $N_{\rm dr}^2 f_{\rm drid}^2$ which can be derived as
\begin{align}
    \label{eq:nifi_def}
    n_{\rm drid} = 1 + \dfrac{\ln \Ndr^2 {\cal P}_{\mathcal{II}}^{(1)} - \ln \Ndr^2 {\cal P}_{\mathcal{II}}^{(2)}}{\ln k_1 - \ln k_2},&\qquad 
    \Ndr^2 A_{\rm drid} = \Ndr^2 {\cal P}_{\mathcal{II}}^{(1)} \exp\left[(n_{\rm drid} -1 )\ln\left(k_\ast \over k_1\right)\right]\;,\nonumber\\
    \Ndr^2 f_{\rm drid}^2 =& \Ndr^2 \dfrac{A_{\rm drid}}{A_s} = \Ndr^2 {{\cal P}_{\mathcal{II}}^{(1)} \over {\cal P}_{\mathcal{RR}}^{(1)}} \left( k_\ast \over k_1\right)^{(n_{\rm drid} - n_s)}\;.
\end{align}
Following Ref.~\cite{Planck:2018jri}, we have chosen $k_1 = 0.002 ~{\rm Mpc}^{-1}$ , $k_2 = 0.1 ~{\rm Mpc}^{-1}$ and $k_\ast = 0.05~{\rm Mpc}^{-1}$. 
\begin{figure}
    \centering
    \includegraphics[width=0.7\linewidth]{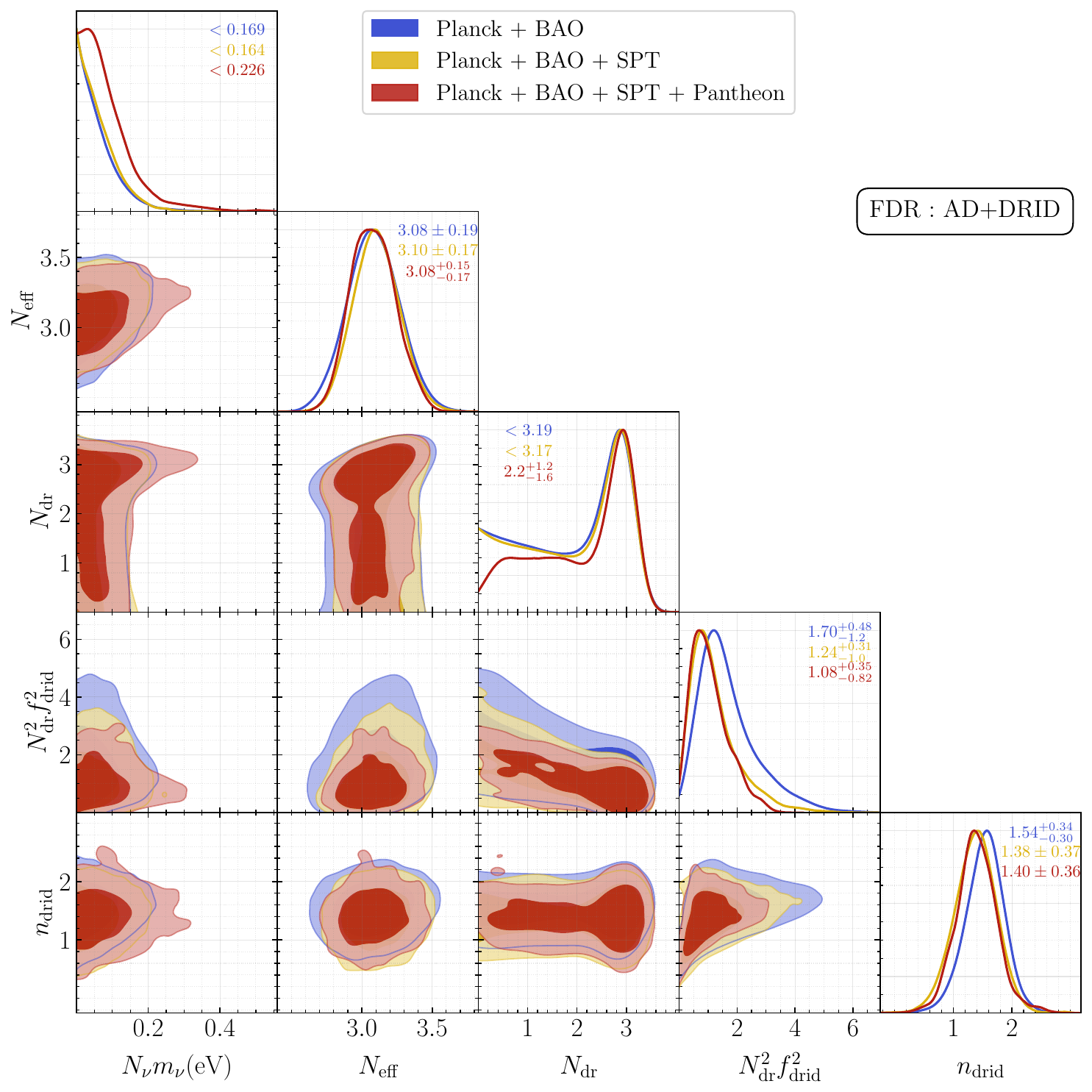}
    \caption{Triangle plots for parameters FDR carrying both adiabatic and isocurvature perturbations (AD+DRID). Constraints are noted at the upper corners of the 1D contours. The error bars are at $1\sigma$ and the upper limits are mentioned at $95\%$ confidence level (C.L.) }
    \label{fig:addrid-fdr}
\end{figure}
Figure~\ref{fig:addrid-fdr} shows the constraint on AD+DRID for the FDR scenario. Similar to the adiabatic perturbation case, even in the presence of DRID, $\Ndr$ constitutes most of the $\Neff$.  However, compared to the adiabatic only case in figure~\ref{fig:adia-fdr-cdr}, the MCMC sample has more allowed points close to $\Ndr \to 0$. Therefore, the 1D posterior of $\Ndr$ is marginally bi-modal in nature. The data prefers a strongly blue-tilted DRID spectrum with $n_{\rm drid}  \approx 1.5$. The scaled isocurvature amplitude fraction $\Ndr^2f_{\rm drid}^2$ shows a slight preference for non-zero values which is consistent with the slight improvement of $\chi^2$ over $\Lambda$CDM. The value of the $\Neff$ is slightly larger than the adiabatic only case. The presence of blue-tilted isocurvature enhances the CMB anisotropy spectrum at high multipole which is compensated by enhanced damping due to the increase $\Neff$. The constraints for all other parameters  and the $\Delta \chi^2 $ compared to $\Lambda$CDM are shown in table~\ref{tab:fdr-cdr-drid}.

\begin{figure}
    \centering
    \includegraphics[width=0.7\linewidth]{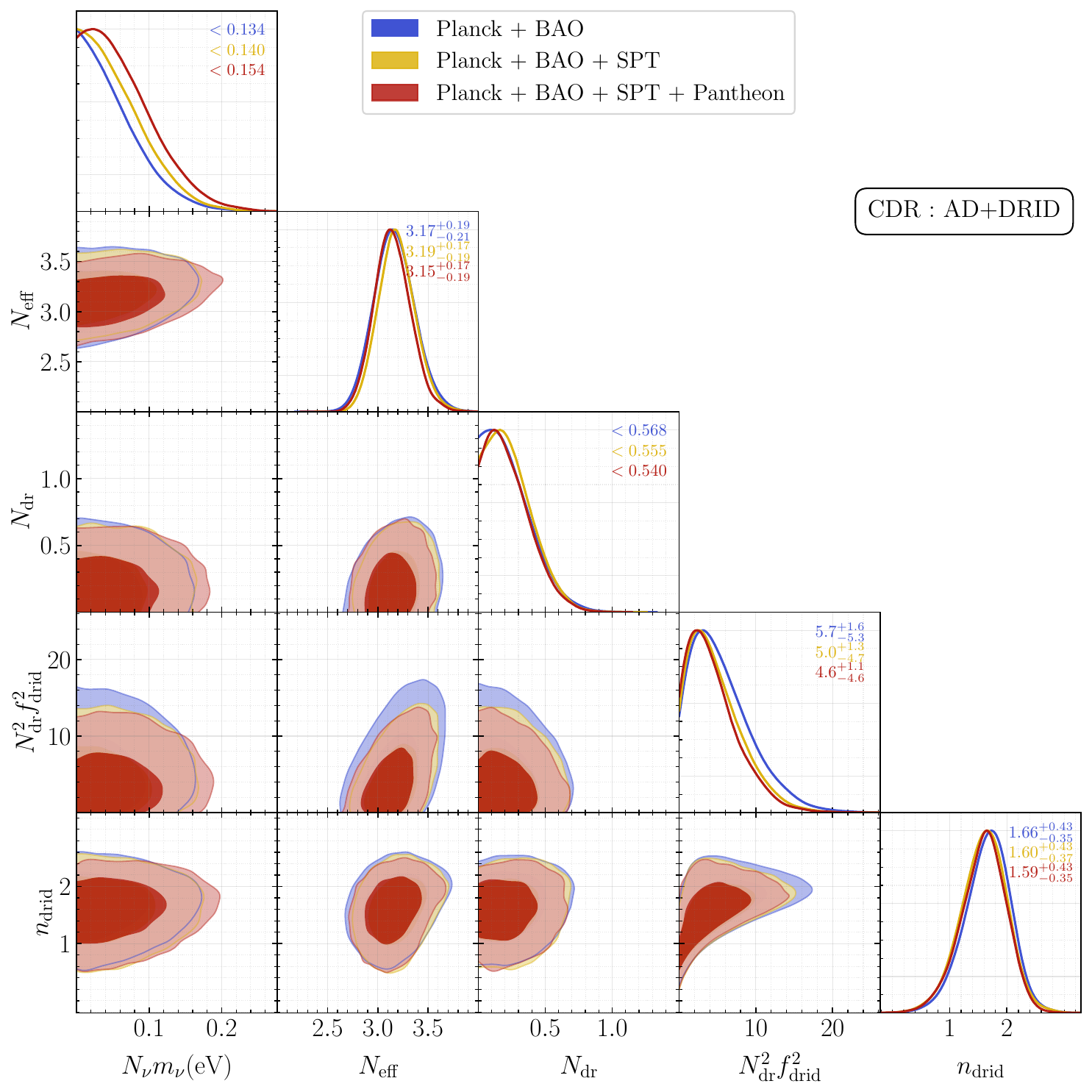}
    \caption{Triangle plots for parameters CDR carrying both adiabatic and isocurvature perturbations. Constraints are noted at the upper corners of the 1D contours. The error bars are at $1\sigma$ and the upper limits are mentioned at $95\%$ confidence level (C.L.) }
    \label{fig:addrid-cdr}
\end{figure}

\begin{table}[]
    \centering
\resizebox{\textwidth}{!}{
    \begin{tabular}{|c||c|c||c|c||c|c||}
    \hline
         &\multicolumn{2}{c||}{Planck + BAO} & \multicolumn{2}{c||}{Planck + BAO + SPT}  & \multicolumn{2}{c||}{Planck + BAO + SPT + PAN} \\
         \hline
         & FDR & CDR & FDR & CDR & FDR & CDR \\
        \hline
$ 10^2 \omega_{b}$ & $2.250\pm 0.020$ & $2.255\pm 0.021$ & $2.244\pm 0.018$ & $2.253\pm 0.019$ & $2.241\pm 0.017$ & $2.247\pm 0.019$\\
\hline
$ \omega{}_{\rm cdm }$ & $0.1199\pm 0.0030$ & $0.1220^{+0.0032}_{-0.0037}$ & $0.1203\pm 0.0027$ & $0.1223^{+0.0028}_{-0.0033}$ & $0.1200^{+0.0023}_{-0.0027}$ & $0.1219^{+0.0028}_{-0.0033}$\\
\hline
$ 100\theta{}_{s }$ & $1.04243\pm 0.00053$ & $1.04259^{+0.00060}_{-0.00074}$ & $1.04216\pm 0.00045$ & $1.04248^{+0.00057}_{-0.00071}$ & $1.04222^{+0.00043}_{-0.00048}$ & $1.04252^{+0.00056}_{-0.00073}$\\
\hline
$ 10^{10}{\cal P}_{\mathcal{RR}}^{(1)}$ & $23.44\pm 0.51$ & $23.54\pm 0.51$ & $23.27\pm 0.48$ & $23.41\pm 0.48$ & $23.35\pm 0.48$ & $23.51\pm 0.48$\\
\hline
$ 10^{10}{\cal P}_{\mathcal{RR}}^{(2)}$ & $20.38\pm 0.41$ & $20.21\pm 0.44$ & $20.40\pm 0.39$ & $20.20\pm 0.43$ & $20.35^{+0.36}_{-0.41}$ & $20.15\pm 0.43$\\
\hline
$ 10^{10}{\cal P}_{\mathcal{II}}^{(1)}N_{\rm dr}^2$ & $< 8.82$ & $< 18.7$ & $< 10.4$ & $< 20.1$ & $< 9.16$ & $< 18.8$\\
\hline
$ 10^{10}{\cal P}_{\mathcal{II}}^{(2)}N_{\rm dr}^2$ & $55^{+20}_{-40}$ & $< 257$ & $37^{+7}_{-30}$ & $< 215$ & $32^{+9}_{-30}$ & $< 196$\\
\hline
$ N_{\rm dr}$ & $< 2.80$ & $< 0.302$ & $< 2.83$ & $< 0.309$ & $2.2^{+1.2}_{-1.6}$ & $< 0.296$\\
\hline
$ N_\nu$ & $< 1.66$ & $2.92\pm 0.24$ & $< 1.68$ & $2.94\pm 0.22$ & $< 1.35$ & $2.91^{+0.24}_{-0.21}$\\
\hline
$ m_\nu ({\rm eV})$ & $< 0.219$ & $< 0.0222$ & $< 0.235$ & $< 0.0244$ & $< 0.613$ & $< 0.0278$\\
\hline
\hline
$ N_{\rm eff}$ & $3.08\pm 0.19$ & $3.17^{+0.19}_{-0.21}$ & $3.10\pm 0.17$ & $3.19^{+0.17}_{-0.19}$ & $3.08^{+0.15}_{-0.17}$ & $3.15^{+0.17}_{-0.19}$\\
\hline
$ N_\nu m_\nu (\rm eV)$ & $< 0.0729$ & $< 0.0646$ & $< 0.0762$ & $< 0.0719$ & $< 0.0997$ & $< 0.0814$\\
\hline
$ H_0 ({\rm km/s/Mpc})$ & $68.2\pm 1.2$ & $68.6\pm 1.3$ & $68.1\pm 1.1$ & $68.7\pm 1.2$ & $67.8\pm 1.0$ & $68.2\pm 1.1$\\
\hline
$ \sigma_8$ & $0.808^{+0.014}_{-0.011}$ & $0.811^{+0.012}_{-0.011}$ & $0.808^{+0.013}_{-0.011}$ & $0.810^{+0.012}_{-0.010}$ & $0.804^{+0.014}_{-0.010}$ & $0.808^{+0.012}_{-0.010}$\\
\hline
$ 10^{-9}A_{s }$ & $2.089^{+0.033}_{-0.037}$ & $2.076\pm 0.038$ & $2.088\pm 0.034$ & $2.073\pm 0.037$ & $2.085^{+0.032}_{-0.036}$ & $2.071\pm 0.037$\\
\hline
$ n_{s }$ & $0.9642\pm 0.0076$ & $0.9609\pm 0.0079$ & $0.9664\pm 0.0071$ & $0.9623\pm 0.0076$ & $0.9648\pm 0.0067$ & $0.9606\pm 0.0076$\\
\hline
$ N_{\rm dr}^2 f^2_{\rm drid}$ & $1.70^{+0.48}_{-1.2}$ & $5.7^{+1.6}_{-5.3}$ & $1.24^{+0.31}_{-1.0}$ & $5.0^{+1.3}_{-4.7}$ & $1.08^{+0.35}_{-0.82}$ & $4.6^{+1.1}_{-4.6}$\\
\hline
$ n_{\rm drid}$ & $1.54^{+0.34}_{-0.30}$ & $1.66^{+0.43}_{-0.35}$ & $1.38\pm 0.37$ & $1.60^{+0.43}_{-0.37}$ & $1.40\pm 0.36$ & $1.59^{+0.43}_{-0.35}$\\
\hline
$\Delta \chi^2$ & $-5.0$ & $-1.0$ & $-0.3$ & $-0.2$ & $-2.4$ & $0.9$\\
\hline
AIC & $5.0$ & $9.0$ & $9.7$ & $9.8$ & $7.6$ & $10.9$\\
\hline
    \end{tabular}
}
    \caption{Marginalized constraints in the primary and derived cosmological parameters (below double line) for the FDR and CDR case carrying of AD+DRID perturbations for all datasets. The error bars are at $1\sigma$ and upper limits are at $68\%$ C.L. }
    \label{tab:fdr-cdr-drid}
\end{table}
The CDR case is shown in figure~\ref{fig:addrid-cdr}. In this case, $N_\nu$ contributes the most to $\Neff$, similar to the adiabatic only case. The presence of additional DRID perturbations cannot offset the effects of CDR adiabatic perturbations on CMB. Thus, in this case, DR is the subdominant component of the total $\Neff$.
The presence of CDR isocurvature induces some additional features which can be seen in the MCMC constraints. Figure~\ref{fig:addrid-cdr} shows that all datasets advocate for a stronger blue-titled isocurvature with a tilt larger than the FDR case. 
The value of the scaled isocurvature amplitude fraction $(\Ndr f_{\rm drid})$ is also larger in the CDR case. In the presence of DRID, the power spectra for the CDR are suppressed compared to FDR due to suppressed free-streaming~\cite{Ghosh:2021axu}. The value of total $\Neff$ is also larger in this case. A strongly blue-tilted isocurvature and suppressed free-streaming both enhance the CMB tail, which allows for larger $\Neff$ to compensate those effects via additional Silk damping. 
Note that the $68\%$ C.L. constraint on $N_\nu m_\nu $ is slightly weaker in the presence of DRID.
\begin{figure}
    \centering
    \includegraphics[width=0.7\linewidth]{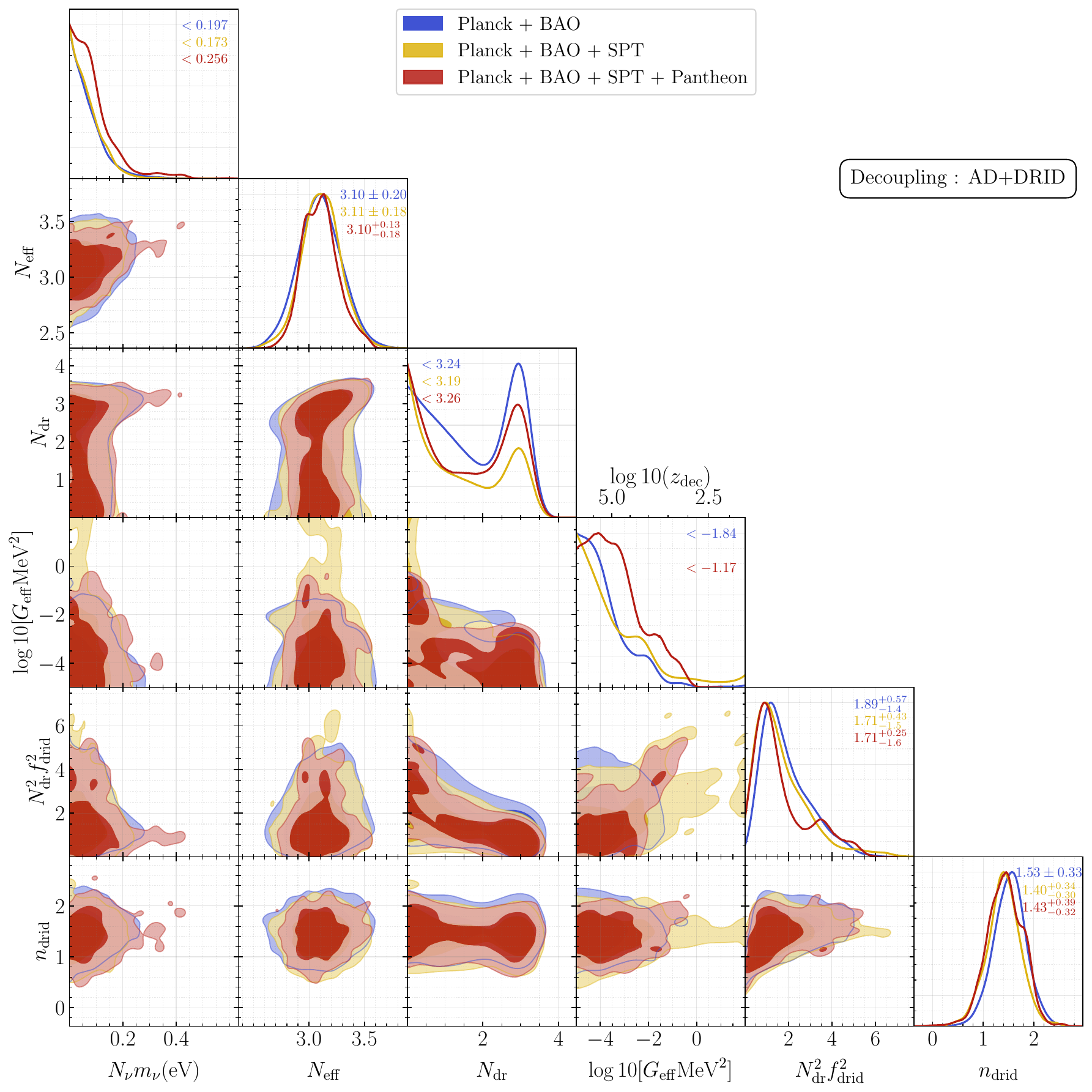}
    \caption{Triangle plots for parameters for decoupling DR carrying both adiabatic and isocurvature perturbations. Constraints are noted at the upper corners of the 1D contours. The error bars are at $1\sigma$ and the upper limits are mentioned at $95\%$ confidence level (C.L.) }
    \label{fig:addrid-decoup}
\end{figure}

\begin{figure}
    \centering
    \includegraphics[width=0.7\linewidth]{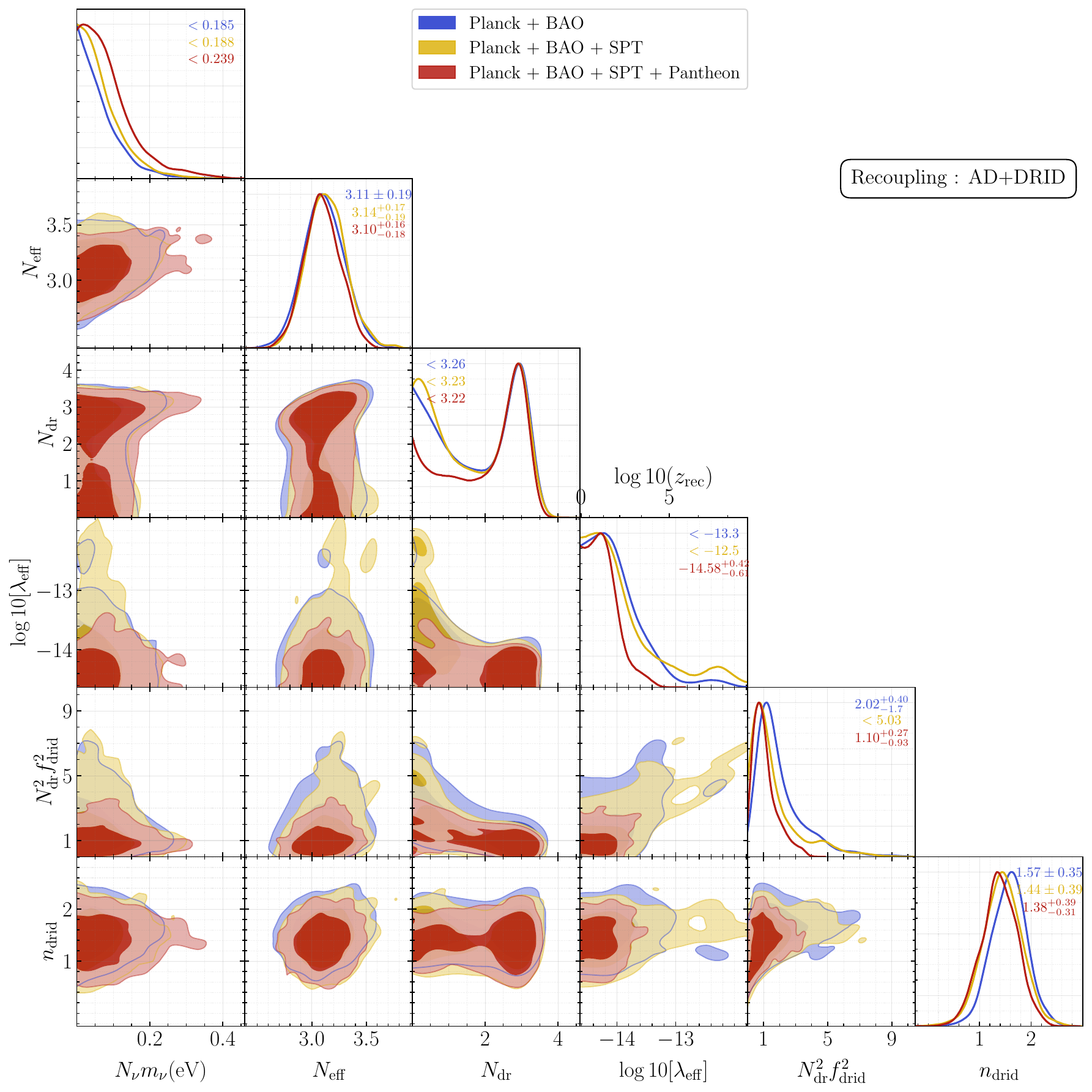}
    \caption{Triangle plots for parameters for recoupling DR carrying both adiabatic and isocurvature perturbations. Constraints are noted at the upper corners of the 1D contours. The error bars are at $1\sigma$ and the upper limits are mentioned at $95\%$ confidence level (C.L.). }
    \label{fig:addrid-recoup}
\end{figure}

Next, we discuss the effects of adding self-interactions in the DR sector and the resulting effects on the isocurvature constraints. Figures~\ref{fig:addrid-decoup} and~\ref{fig:addrid-recoup} show the constraint on the relevant quantities for the decoupling and recoupling cases, respectively. We don't find any evidence for interacting DR in either case. The constraints in both decoupling and recoupling scenarios are in between those of the CDR and FDR scenarios. In the top $x$-axis of the 1D posterior of $\lgeff$ and $\lleff$ in figures~\ref{fig:addrid-decoup} and \ref{fig:addrid-recoup}, we show the corresponding values of $z_{\rm dec}$ and $z_{\rm rec}$, respectively. We constrain $z_{\dec} \gtrsim 10^5$ and $z_{\rm rec} \lesssim 300$, which is similar and slightly stronger than the respective constraints in the adiabatic-only case. Note that these constraints are marginalized over all values of $\Ndr$. For a smaller value of $\Ndr$, much stronger interactions are allowed, which can be seen for the corresponding 2D contours. Specifically for $\Ndr \approx 1$ the $95\%$ C.L. limit on the interaction strengths is approximately $\lgeff \lesssim -2 ~(z_{\rm dec } \gtrsim 10^4 )$ and $\lleff \lesssim -13.4~ (z_{\rm rec } \lesssim 10^4 )$ for the decoupling and recoupling cases respectively. Thus, for $\Ndr \approx 1$, interactions can keep DR coupled around matter-radiation equality. For the decoupling case, this result is in accordance with studies of flavor-specific neutrino or DR self-interaction with adiabatic perturbations~\cite{Das:2020xke,Brinckmann:2020bcn,Brinckmann:2022ajr,RoyChoudhury:2022rva,Das:2023npl}. Note that the $\Ndr$ posteriors in the presence of interaction show strong bimodality, which results in much relaxed $95\%$ C.L. constraints for both $N_\nu$ and $\Ndr$. 

\begin{table}[]
    \centering
\resizebox{\textwidth}{!}{
    \begin{tabular}{|c||c|c||c|c||c|c||}
    \hline
         &\multicolumn{2}{c||}{Planck + BAO} & \multicolumn{2}{c||}{Planck + BAO + SPT}  & \multicolumn{2}{c||}{Planck + BAO + SPT + PAN} \\
         \hline
         & Decoupling & Recoupling & Decoupling & Recoupling & Decoupling & Recoupling \\
        \hline
$ 10^2 \omega_{b}$ & $2.250\pm 0.020$ & $2.252\pm 0.020$ & $2.245^{+0.017}_{-0.020}$ & $2.247\pm 0.019$ & $2.243^{+0.016}_{-0.019}$ & $2.242\pm 0.018$\\
\hline
$ \omega{}_{\rm cdm }$ & $0.1201\pm 0.0032$ & $0.1204\pm 0.0031$ & $0.1205\pm 0.0029$ & $0.1209\pm 0.0031$ & $0.1205^{+0.0025}_{-0.0023}$ & $0.1204\pm 0.0026$\\
\hline
$ 100\theta{}_{s }$ & $1.04260^{+0.00044}_{-0.00073}$ & $1.04251\pm 0.00056$ & $1.04234^{+0.00041}_{-0.00072}$ & $1.04226\pm 0.00051$ & $1.04230^{+0.00039}_{-0.00048}$ & $1.04220^{+0.00044}_{-0.00051}$\\
\hline
$ 10^{10}{\cal P}_{\mathcal{RR}}^{(1)}$ & $23.54\pm 0.56$ & $23.32\pm 0.53$ & $23.37\pm 0.52$ & $23.18\pm 0.52$ & $23.30\pm 0.47$ & $23.27\pm 0.48$\\
\hline
$ 10^{10}{\cal P}_{\mathcal{RR}}^{(2)}$ & $20.28^{+0.46}_{-0.40}$ & $20.38\pm 0.41$ & $20.30^{+0.47}_{-0.41}$ & $20.40\pm 0.39$ & $20.32^{+0.44}_{-0.52}$ & $20.38^{+0.37}_{-0.42}$\\
\hline
$ 10^{10}{\cal P}_{\mathcal{II}}^{(1)}N_{\rm dr}^2$ & $< 9.71$ & $< 9.10$ & $< 12.7$ & $< 10.8$ & $< 10.6$ & $< 9.37$\\
\hline
$ 10^{10}{\cal P}_{\mathcal{II}}^{(2)}N_{\rm dr}^2$ & $60^{+20}_{-50}$ & $67^{+10}_{-60}$ & $< 58.6$ & $< 53.6$ & $< 56.4$ & $32^{+8}_{-30}$\\
\hline
$ \log10[G_{\rm eff} ~{\rm MeV}^2]$ & $< -3.59$ & $--$ & $< -2.99$ & $--$ & $< -2.96$ & $--$\\
\hline
$ \log10[\lambda_{\rm eff}]$  & $--$ & $< -14.1$ & $--$ & $< -14.1$ & $--$ & $-14.58^{+0.42}_{-0.61}$\\
\hline
$ N_{\rm dr}$ & $< 2.76$ & $< 2.82$ & $< 2.40$ & $< 2.79$ & $< 2.72$ & $< 2.88$\\
\hline
$ N_\nu$ & $< 2.09$ & $< 2.06$ & $< 2.45$ & $< 2.23$ & $< 2.18$ & $< 1.43$\\
\hline
$ m_\nu ({\rm eV})$ & $< 0.202$ & $< 0.256$ & $< 0.0911$ & $< 0.193$ & $< 0.209$ & $< 0.488$\\
\hline
\hline
$ N_{\rm eff}$ & $3.10\pm 0.20$ & $3.11\pm 0.19$ & $3.11\pm 0.18$ & $3.14^{+0.17}_{-0.19}$ & $3.10^{+0.13}_{-0.18}$ & $3.10^{+0.16}_{-0.18}$\\
\hline
$ N_\nu m_\nu (\rm eV)$ & $< 0.0842$ & $< 0.0782$ & $< 0.0829$ & $< 0.0829$ & $< 0.0977$ & $< 0.103$\\
\hline
$ H_0 ({\rm km/s/Mpc})$ & $68.3\pm 1.2$ & $68.4\pm 1.2$ & $68.2\pm 1.1$ & $68.3^{+1.1}_{-1.2}$ & $67.89^{+0.84}_{-1.0}$ & $67.9\pm 1.1$\\
\hline
$ \sigma_8$ & $0.809^{+0.014}_{-0.012}$ & $0.809^{+0.014}_{-0.011}$ & $0.810^{+0.014}_{-0.011}$ & $0.809^{+0.014}_{-0.010}$ & $0.804^{+0.013}_{-0.010}$ & $0.805^{+0.015}_{-0.011}$\\
\hline
$ 10^{-9}A_{s }$ & $2.082^{+0.039}_{-0.035}$ & $2.087\pm 0.036$ & $2.081^{+0.039}_{-0.035}$ & $2.086\pm 0.034$ & $2.081^{+0.034}_{-0.040}$ & $2.086^{+0.031}_{-0.037}$\\
\hline
$ n_{s }$ & $0.9619^{+0.0095}_{-0.0071}$ & $0.9656\pm 0.0077$ & $0.9639^{+0.0087}_{-0.0077}$ & $0.9674\pm 0.0075$ & $0.9650^{+0.0067}_{-0.0075}$ & $0.9661\pm 0.0070$\\
\hline
$ N_{\rm dr}^2 f^2_{\rm drid}$ & $1.89^{+0.57}_{-1.4}$ & $2.02^{+0.40}_{-1.7}$ & $1.71^{+0.43}_{-1.5}$ & $< 1.76$ & $1.71^{+0.25}_{-1.6}$ & $1.10^{+0.27}_{-0.93}$\\
\hline
$ n_{\rm drid}$ & $1.53\pm 0.33$ & $1.57\pm 0.35$ & $1.40^{+0.34}_{-0.30}$ & $1.44\pm 0.39$ & $1.43^{+0.39}_{-0.32}$ & $1.38^{+0.39}_{-0.31}$\\
\hline
$ \log10 (z_{\rm dec})$ & $5.18^{+0.76}_{-0.25}$ & $--$ & $4.91^{+1.1}_{-0.67}$ & $--$ & $4.90^{+0.95}_{-0.41}$ & $--$\\
\hline
$ \log10 (z_{\rm rec})$ & $--$ & $< 2.44$ & $--$ & $< 2.47$ & $--$ & $-1.1\pm 3.5$\\
\hline
$\Delta \chi^2$ & $-5.2$ & $-5.1$ & $-1.5$ & $-2.1$ & $-2.8$ & $-2.9$\\
\hline
AIC & $6.8$ & $6.9$ & $11.0$ & $12.1$ & $9.2$ & $9.1$\\
\hline
    \end{tabular}
}
    \caption{Marginalized constraints in the primary and derived cosmological parameters (below double line) for the decoupling and recoupling cases carrying of AD+DRID perturbations for all datasets. The error bars are at $1\sigma$ and upper limits are at $68\%$ C.L. }
    \label{tab:fdr-cdr-drid}
\end{table}

\section{ Implications for the $H_0$ Tension}\label{sec:H0}

In this section, we study the implications of DR interactions and the presence of isocurvature initial conditions for the Hubble tension. The discrepancy between the measured value of $H_0$ using the distance ladder (from the SH0ES collaboration) and the inferred value of the $H_0$ from CMB observations (assuming the $\Lambda$CDM model) currently stands at approximately $6 \sigma$~\cite{Riess:2025chq}.\footnote{Note that, the determination of $H_0 = 70.39 \pm 1.22 ({\rm stat}) \pm
1.33 ({\rm sys}) \pm 0.70 (\sigma{\rm SN})$ km/s/Mpc  by the Chicago-Carnegie Hubble Program (CCHP) using the tip of the red giant branch and J-Region Asymptotic Giant Branch~\cite{Freedman:2024eph} is significantly smaller with a larger error bar than the measurement of the SH0ES collaboration: $H_0 = 73.18 \pm 0.88$ km/s/Mpc~\cite{Riess:2025chq}. The CCHP value suggests a much milder tension/discrepency $(\approx 2 \sigma)$ with the CMB measurement of $H_0$} Several beyond the $\Lambda$CDM extensions have been proposed to ameliorate this tension. However, only a handful of scenarios are able to reduce the tension substantially~(for reviews see Refs.~\cite{Schoneberg:2021qvd,Knox:2019rjx,DiValentino:2021izs,Khalife:2023qbu}). In this section, we investigate whether the DR isocurvature and DR interactions are able to mitigate the $H_0$ tension. To quantify the tension and model comparison, we used two metrics used in Ref.~\cite{Schoneberg:2021qvd},
\begin{itemize}
    \item {\bf Gaussian Tension (GT)} defined as,
    \begin{equation}
        \rm{GT} = \dfrac{\bar{H}_{0,\mathcal{D}} - \bar{H}_{0,{\rm SH0ES}}}{\sqrt{\sigma_\mathcal{D}^2 + \sigma_{\rm SH0ES}^2  }}\;,
    \end{equation}
    where $\bar{H}_0$ and $\sigma$ are the mean value and the uncertainty of the $H_0$ measurement.  $\mathcal{D}$ refers to the dataset that we set to be the `Planck + BAO + SPT + PAN'. GT essentially measures with all the available datasets (except SH0ES) at hand, how far we can reduce the $H_0$ tension. Notably, $\mathcal{D}$ doesn't include the SH0ES data and is not biased to generate a larger $H_0$. We used the latest determination of the Hubble constant from the SH0ES collaboration for this metric $H_0 = 73.18 \pm 0.88$ km/s/Mpc~\cite{Riess:2025chq}.   
    \item The Akaike Information Criterion (AIC) defined earlier in Eq.~\eqref{eq:AIC} also serves as a model comparison, which penalizes models containing lots of new parameters that can lead to overfitting. However, the addition of a single physical feature in a model can naturally bring in multiple parameters. The dark radiation isocurvature, for example, needs two additional parameters (magnitude and tilt) to characterize the isocurvature spectrum (four if correlation with adiabatic mode is non-zero, which is not considered here). Consequently, such features may lead to a larger AIC. We have computed the AIC for all the models for the datasets in the parameter tables.
\end{itemize}

\begin{table}[]
    \centering
\resizebox{\textwidth}{!}{
    \begin{tabular}{|c||c||c|c||c|c||c|c||c|c|}
    \hline
         &\multicolumn{9}{c|}{Planck + BAO + SPT + PAN} \\
         \hline
        &$\Lambda$CDM &\multicolumn{2}{c||}{Decoupling} & \multicolumn{2}{c||}{Recoupling}
         & \multicolumn{2}{c||}{FDR} &
         \multicolumn{2}{c|}{CDR}\\
        \hline
        &AD& AD+DRID & AD & AD+DRID & AD & AD+DRID & AD & AD+DRID & AD \\
        \hline
$ N_{\rm eff}$ &  $3.044$ & $3.10^{+0.13}_{-0.18}$ & $3.01^{+0.14}_{-0.16}$ & $3.10^{+0.16}_{-0.18}$ & $3.00\pm 0.16$ & $3.08^{+0.15}_{-0.17}$ & $2.99\pm 0.15$ & $3.15^{+0.17}_{-0.19}$ & $3.03\pm 0.16$\\
\hline
$ H_0 ({\rm km/s/Mpc})$ & $67.8^{+0.38}_{-0.39}$ &$67.89^{+0.84}_{-1.0}$ & $67.26^{+0.87}_{-1.0}$ & $67.9\pm 1.1$ & $67.20^{+0.94}_{-1.1}$ & $67.8\pm 1.0$ & $67.15\pm 0.97$ & $68.2\pm 1.1$ & $67.5\pm 1.0$\\
\hline
GT & $5.6$ & $4.15$ & $4.6$ & $3.75$ & $4.43$ & $4.04$ & $4.6$ & $3.54$ & $4.26$\\
\hline

\end{tabular}
}
    \caption{$\Neff$, $H_0$ and Gaussian Tension (GT) for all the DR cases for Planck + BAO + SPT + PAN dataset. The GT are calculated using the latest SH0ES data $H_0 = 73.18 \pm 0.88$ km/s/Mpc~\cite{Riess:2025chq}. Presence of DR interactions and DR isocurvature (DRID) allows for a larger $\Neff$, which gives a higher $H_0,$ reducing the GT with SH0ES.  }
    \label{tab:GT}
\end{table}

Table.~\ref{tab:GT} shows the values of GT along with values of $\Neff$ and $H_0$ for all the cases studied here. For the $\Lambda$CDM model, the GT stands at $5.6\sigma$. Adding interactions alone in the DR reduces the tension to $\approx 4.2\sigma$ (for the CDR case). Note that even in the FDR case, where there are no interactions, the tension reduces to $4.6\sigma$. Since our models come with additional free parameters (especially $\Neff$), there is increased degeneracy between parameters. Additional degeneracies lead to an increase in the error bar in the parameter constraints, such as $H_0$, which also reduces the tension. Note that the central value of $H_0$ does not increase (rather decreases) for the AD runs compared to $\Lambda$CDM. This is an artifact of the positive correlation between $\Neff$ and $H_0$. For the $\Lambda$CDM case, $\Neff$ is kept fixed to $3.044$, whereas it is varied (through $\Ndr$ and $N_\nu$) for DR cases. When $\Neff$ is allowed to vary, the CMB data prefers a smaller value of $\Neff$ compared to the standard value of $3.044$~\cite{Planck:2018vyg}. This is clearly visible in the FDR (AD) case, where $\Neff \approx 2.99 $. Thus, a smaller $\Neff$ leads to a smaller central value of $H_0$, but increased degeneracy results in a larger error bar. DR interactions increase the value of $\Neff$, hence $H_0$, reducing the tension even more. 
The addition of DR isocurvature further increases the $\Neff$ due to the strong blue tilt, as discussed in the previous section.
This, in turn, increases the central value of $H_0$. 
On the other hand, the error bar increase due to the addition of DRID is very marginal. 
Thus, due to the shift towards higher $H_0$, in the DRID case, GT is reduced significantly. 
The best case scenario is CDR with DRID, for which the GT is reduced to $\approx 3.5\sigma$.

\begin{figure}
    \centering
    \includegraphics[width=0.8\linewidth]{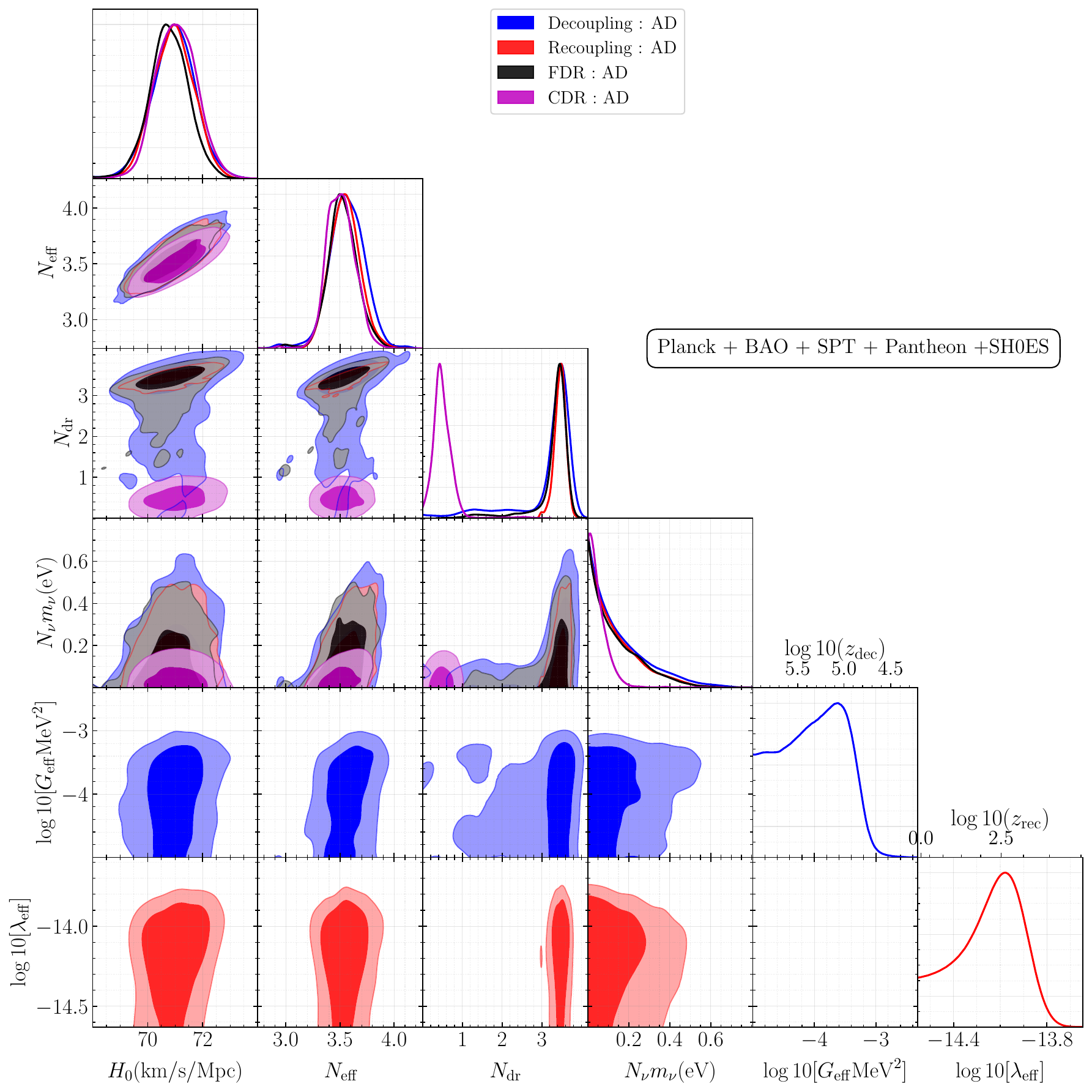}
    \caption{Triangle plots for parameters all DR interaction cases with only adiabatic perturbations for Planck + BAO + SPT + PAN + SH0ES dataset.
    }
    \label{fig:tri_sh0es_adi}
\end{figure}

\begin{figure}
    \centering
    \includegraphics[width=0.8\linewidth]{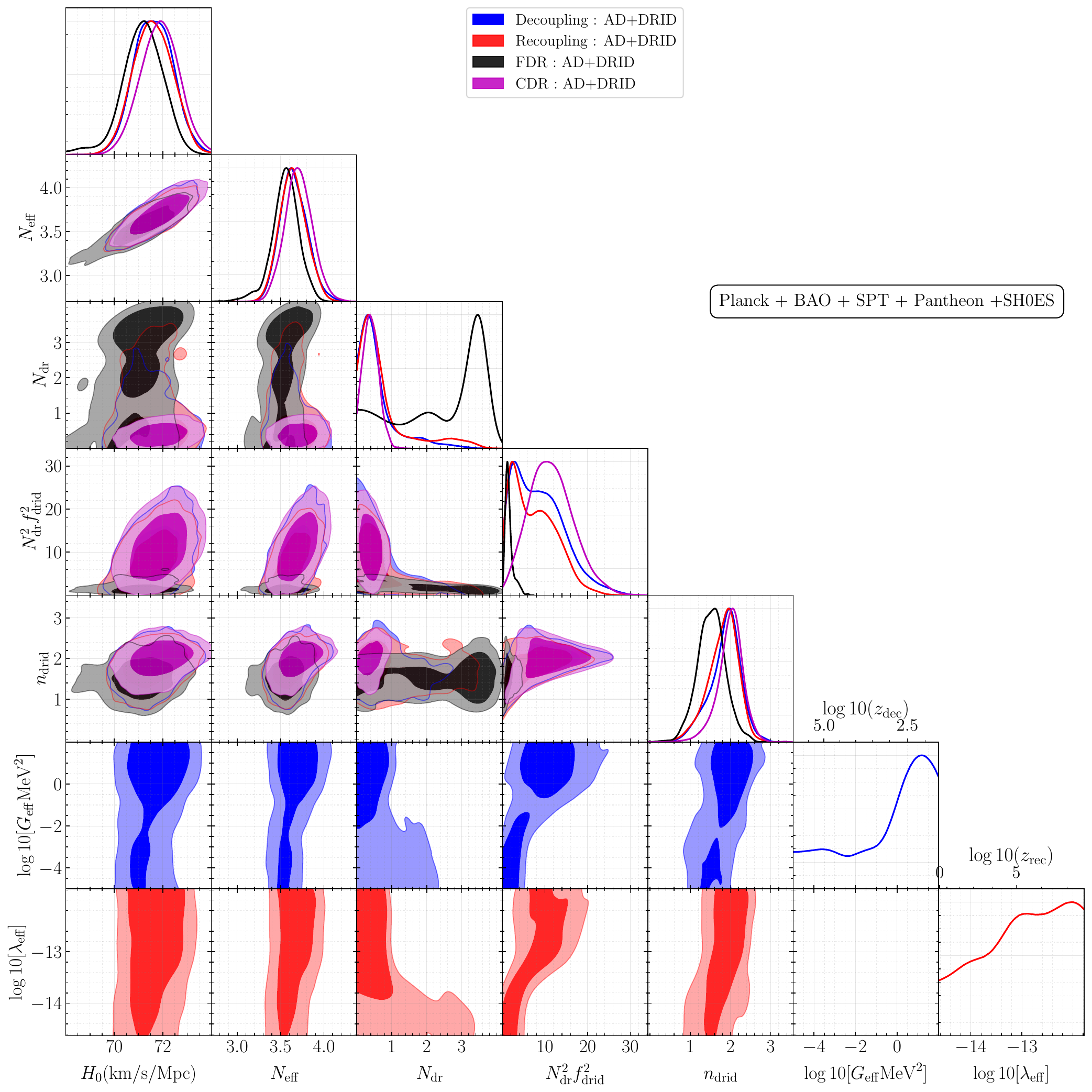}
    \caption{Triangle plots for parameters all DR interaction cases with both adiabatic and DRID perturbations for Planck + BAO + SPT + PAN + SH0ES dataset.
    }
    \label{fig:tri_sh0es_adi_drid}
\end{figure}

Thus, although DRID and DR interactions help to alleviate the tension, they fail to resolve the tension. While there is an upward shift in $H_0$, the shift is rather small and is unable to move the $H_0$ substantially closer to the SH0ES value. Finally, for completeness, we also show the parameter constraints of the models for the Planck + BAO + SPT + PAN + SH0ES dataset in figures~\ref{fig:tri_sh0es_adi} and \ref{fig:tri_sh0es_adi_drid} and in table~\ref{tab:sh0es}. 

For the combined analysis with adiabatic perturbations, DR comprises most of the $\Neff$ in the FDR, decoupling and recoupling cases. We find no evidence of strong interaction during recombination in the decoupling or recoupling cases. For the recoupling case, the data seem to have a mild preference for the coupling strength corresponding to $z_{\rm rec} \approx 300$. Note that the total $\Neff$ across all cases is substantially higher $( \Neff \approx 3.5)$ which drives the value of the Hubble constant to $H_0 \approx 71$ km/s/Mpc.

In the presence of DRID, DR becomes a subdominant component of the total $\Neff$ both in the decoupling and recoupling cases. The data also support strong interaction which keeps DR strongly coupled during equality and recombination eras. It also accommodates large isocurvature where the mean value of $\Ndr^2 f_{\rm drid}^2$ increases compared to the analysis without the SH0ES dataset in the interacting DR cases. The DRID is also strongly blue-tilted with  $n_{\rm drid} \approx 2$. The strong coupling and the large isocurvature in DR sector push $\Neff$ to a larger value $\approx 3.6$ resulting in a $H_0 \approx 71.5$ km/s/Mpc.

\begin{table}[]
    \centering
\resizebox{\textwidth}{!}{
    \begin{tabular}{|c||c|c||c|c||c|c||c|c|}
    \hline
         &\multicolumn{8}{c|}{Planck + BAO + SPT + PAN + SH0ES} \\
         \hline
         &\multicolumn{2}{c||}{Decoupling} & \multicolumn{2}{c||}{Recoupling}
         & \multicolumn{2}{c||}{FDR} &
         \multicolumn{2}{c|}{CDR}\\
        \hline
        & AD+DRID & AD & AD+DRID & AD & AD+DRID & AD & AD+DRID & AD \\
        \hline
$ 10^2 \omega_{b}$ & $2.291^{+0.015}_{-0.017}$ & $2.277\pm 0.018$ & $2.292\pm 0.016$ & $2.279\pm 0.015$ & $2.286\pm 0.017$ & $2.279\pm 0.016$ & $2.295\pm 0.015$ & $2.284\pm 0.016$\\
\hline
$ \omega{}_{\rm cdm }$ & $0.1295^{+0.0029}_{-0.0035}$ & $0.1261^{+0.0028}_{-0.0023}$ & $0.1292^{+0.0029}_{-0.0032}$ & $0.1260\pm 0.0023$ & $0.1263^{+0.0027}_{-0.0021}$ & $0.1256\pm 0.0024$ & $0.1308\pm 0.0031$ & $0.1270\pm 0.0025$\\
\hline
$ 100\theta{}_{s }$ & $1.04197^{+0.00054}_{-0.00087}$ & $1.04102^{+0.00035}_{-0.00040}$ & $1.04195^{+0.00051}_{-0.00079}$ & $1.04115\pm 0.00041$ & $1.04149\pm 0.00044$ & $1.04101\pm 0.00039$ & $1.04206^{+0.00066}_{-0.00079}$ & $1.04243^{+0.00058}_{-0.00085}$\\
\hline
$ 10^{10}{\cal P}_{\mathcal{RR}}^{(1)}$ & $22.93\pm 0.51$ & $22.86\pm 0.45$ & $22.74\pm 0.53$ & $22.48\pm 0.46$ & $22.60^{+0.43}_{-0.50}$ & $22.74^{+0.38}_{-0.45}$ & $23.03\pm 0.49$ & $22.93^{+0.36}_{-0.41}$\\
\hline
$ 10^{10}{\cal P}_{\mathcal{RR}}^{(2)}$ & $20.57\pm 0.49$ & $21.12\pm 0.41$ & $20.67\pm 0.49$ & $21.20^{+0.34}_{-0.39}$ & $20.97\pm 0.41$ & $21.20^{+0.33}_{-0.39}$ & $20.46\pm 0.46$ & $20.49\pm 0.46$\\
\hline
$ 10^{10}{\cal P}_{\mathcal{II}}^{(1)}N_{\rm dr}^2$ & $< 15.2$ & $--$ & $< 15.1$ & $--$ & $< 9.06$ & $--$ & $< 14.6$ & $--$\\
\hline
$ 10^{10}{\cal P}_{\mathcal{II}}^{(2)}N_{\rm dr}^2$ & $< 485$ & $--$ & $< 411$ & $--$ & $52^{+10}_{-40}$ & $--$ & $497^{+200}_{-300}$ & $--$\\
\hline
$ \log10[G_{\rm eff} {\rm MeV}^2]$ & $> -1.22$ & $-4.04^{+0.69}_{-0.45}$ & $--$ & $--$ & $--$ & $--$ & $--$ & $--$\\
\hline
$ \log10[\lambda_{\rm eff}]$& $--$ & $--$ & $> -13.6$ & $-14.38^{+0.51}_{-0.23}$  & $--$ & $--$ & $--$ & $--$\\
\hline
$ N_{\rm dr}$ & $< 0.615$ & $3.19^{+0.58}_{-0.0099}$ & $< 0.688$ & $3.46^{+0.15}_{-0.13}$ & $< 3.46$ & $3.30^{+0.34}_{-0.056}$ & $0.41^{+0.19}_{-0.25}$ & $0.50^{+0.17}_{-0.24}$\\
\hline
$ N_\nu$ & $3.04^{+0.62}_{-0.15}$ & $< 0.155$ & $2.88^{+0.77}_{-0.11}$ & $< 0.0923$ & $< 1.36$ & $< 0.122$ & $3.30\pm 0.25$ & $3.00^{+0.27}_{-0.20}$\\
\hline
$ m_\nu ({\rm eV})$ & $< 0.0303$ & $---$ & $< 0.0348$ & $---$ & $< 1.54$ & $---$ & $< 0.0284$ & $< 0.0221$\\
\hline
$ N_{\rm eff}$ & $3.65^{+0.14}_{-0.17}$ & $3.56\pm 0.17$ & $3.64^{+0.14}_{-0.16}$ & $3.55\pm 0.14$ & $3.55^{+0.16}_{-0.13}$ & $3.52\pm 0.15$ & $3.71\pm 0.16$ & $3.51^{+0.11}_{-0.14}$\\
\hline
$ N_\nu m_\nu (\rm eV)$ & $< 0.0906$ & $< 0.190$ & $< 0.0956$ & $< 0.169$ & $< 0.142$ & $< 0.171$ & $< 0.0937$ & $< 0.0664$\\
\hline
$ H_0 ({\rm km/s/Mpc})$ & $71.63\pm 0.80$ & $70.89^{+0.88}_{-0.69}$ & $71.60\pm 0.81$ & $70.94\pm 0.73$ & $71.09^{+0.95}_{-0.71}$ & $70.75\pm 0.81$ & $71.89\pm 0.81$ & $71.02\pm 0.88$\\
\hline
$ \sigma_8$ & $0.825^{+0.013}_{-0.0097}$ & $0.820^{+0.022}_{-0.011}$ & $0.823^{+0.013}_{-0.011}$ & $0.822^{+0.019}_{-0.011}$ & $0.817^{+0.020}_{-0.012}$ & $0.820^{+0.020}_{-0.011}$ & $0.823^{+0.013}_{-0.011}$ & $0.821^{+0.012}_{-0.0092}$\\
\hline
$ 10^{-9}A_{s }$ & $2.097\pm 0.041$ & $2.142\pm 0.036$ & $2.102\pm 0.041$ & $2.142^{+0.031}_{-0.035}$ & $2.125\pm 0.036$ & $2.147^{+0.030}_{-0.035}$ & $2.089\pm 0.039$ & $2.090\pm 0.039$\\
\hline
$ n_{s }$ & $0.9723^{+0.0092}_{-0.0082}$ & $0.9797^{+0.0066}_{-0.0057}$ & $0.976^{+0.010}_{-0.0085}$ & $0.9850^{+0.0051}_{-0.0061}$ & $0.9809^{+0.0070}_{-0.0055}$ & $0.9821\pm 0.0057$ & $0.9697^{+0.0087}_{-0.0079}$ & $0.9713^{+0.0071}_{-0.0063}$\\
\hline
$ N_{\rm dr}^2 f^2_{\rm drid}$ & $8.8^{+3.2}_{-7.9}$ & $--$ & $7.6^{+2.9}_{-7.0}$ & $--$ & $1.61^{+0.35}_{-1.1}$ & $--$ & $11.5^{+4.5}_{-5.8}$ & $--$\\
\hline
$ n_{\rm drid}$ & $1.87^{+0.44}_{-0.29}$ & $--$ & $1.81^{+0.42}_{-0.29}$ & $--$ & $1.53\pm 0.36$ & $--$ & $2.00^{+0.33}_{-0.25}$ & $--$\\
\hline
$ \log10 (z_{\rm dec})$ & $3.12^{+0.53}_{-1.5}$ & $5.35^{+0.29}_{-0.45}$ & $4.05^{+3.5}_{-0.84}$ & $0.43^{+3.4}_{-0.44}$ & $--$ & $--$ & $--$ & $--$\\
\hline
$\Delta \chi^2$ & $-17.0$ & $-10.5$ & $-19.2$ & $-12.8$ & $-18.5$ & $-13.2$ & $-17.9$ & $-13.1$\\
\hline
AIC & $-5.0$ & $-2.5$ & $-7.2$ & $-4.8$ & $-8.5$ & $-7.2$ & $-7.9$ & $-7.1$\\
\hline
\hline      
 \end{tabular}
}
    \caption{Marginalized constraints in the primary and derived cosmological parameters (below double line) for all models for the Planck + BAO + SPT + PAN + SH0ES dataset. The error bars are at $1\sigma$ and limits are at $68\%$ C.L. }
    \label{tab:sh0es}
\end{table}

\section{Conclusion}\label{sec:conclusion}

Dark radiation arises in many well-motivated BSM scenarios that can address both conceptual puzzles of the SM and impact tensions between different measurements in the context of the $\Lambda$CDM model. 
In this work, we focus on the `secluded' DR scenario where DR has only gravitational interactions with SM and DM particles. We demonstrate that precision cosmological measurements provide unique tools to probe secluded DR, which is otherwise difficult to test with terrestrial and astrophysical observations. Despite the lack of interactions with other sectors, secluded DR can still have rich phenomenological features from different kinds of self-interactions as well as different types of initial conditions. In this work, we present a comprehensive study of four different types of dark radiation classified by its self-interactions (free-streaming DR, coupled DR, decoupling DR, and recoupling DR) with a combination of two different initial conditions: adiabatic and isocurvature. In addition, we vary the neutrino energy density, the DR energy density, and the SM neutrino masses. We summarize the key physical effects of these properties on the CMB temperature and lensing power spectrum. We modify the \texttt{CLASS} code to accommodate different kinds of DR and derive constraints using MCMC analysis with recent cosmological data. We find no significant preference for physics beyond the $\Lambda$CDM model, but data exhibit interesting interplays between different physical quantities. We also show that all secluded DR scenarios considered in this work cannot resolve the $H_0$ tension. The best case is CDR with both adiabatic and DR isocurvature, which can reduce the Gaussian tension from $5.6\sigma$ to $3.5\,\sigma$.

\section*{Acknowledgments}
We thank Nicolas Fernandez, Marilena LoVerde, Yuhsin Tsai, and Zachary Weiner for useful discussions.
We also thank Yuhsin Tsai and Zachary Weiner for useful feedback on a draft of this work.
This manuscript has been authored in part by Fermi Forward Discovery Group, LLC under Contract No. 89243024CSC000002 with the U.S. Department of Energy, Oﬃce of Science, Oﬃce of High Energy Physics. PD is supported by the National Natural Science Foundation of China (Grants No. T2388102). SG acknowledges support from the NSF under Grant No. PHY-2413016.
SK is supported in part by the National Science Foundation grant PHY-2210498 and the Simons Foundation. This work used the high-performance computing service at the University of Notre Dame, managed by the Center for Research Computing (CRC) (\url{https://crc.nd.edu}). The authors acknowledge the Texas Advanced Computing Center (TACC) at The University of Texas at Austin for providing computational resources that have contributed to the research results reported within this paper (URL: \url{http://www.tacc.utexas.edu}).
This work was also supported in part through the NYU IT High Performance Computing resources, services, and staff expertise.

\appendix
\section{Conversion between Newtonian and Synchronous Gauge}\label{sec:app_gauge}
In this appendix we review the gauge transformations that relate the synchronous gauge with the Newtonian gauge.
Consider a gauge transformation $x \rightarrow \tilde{x}^\mu = x^\mu -\xi^\mu$, where $\xi^\mu = (\alpha, \partial_i\beta)$ contains only scalar gauge transformation parameters $\alpha$ and $\beta$.
By choosing $\alpha$ and $\beta$ appropriately, one can relate the two gauges.
We first write the general scalar metric perturbations,
\es{eq:gen_metric_2}{
    \D s^2 = a^2\left((-1-2\phi)\D\eta^2 + 2 B_{,i} \D \eta \D x^i + (\delta_{ij} -2\psi \delta_{ij}+2E_{,ij})\D x^i \D x^j\right),
}
Under the scalar gauge transformation perturbation, the various metric components and the perturbation in energy density transform as,
\es{}{
\tilde{\phi} = \phi + {\cal H}\alpha + \alpha',~~\tilde{\psi} = \psi - {\cal H}\alpha,~~\tilde{B} = B - \alpha +\beta',~~\tilde{E} = E+\beta,~~\tilde{\delta\rho} = \delta\rho + \rho'\alpha.
}
Denoting the synchronous gauge quantities with a tilde and Newtonian gauge quantities without it, we choose $\alpha$ and $\beta$ such that $\tilde{\phi}=\tilde{B}=0$.
Given $B=0$ in the Newtonian gauge, we require $\alpha = \beta'$.
Since $E=0$, we also get $\tilde{E} = \beta$.
Now using the notation in Eq.~\eqref{eq:synch} for the synchronous gauge, 
\es{}{
\tilde{\psi} = \eta, ~~\tilde{E} = -{1\over 2 k^2}(h+6\eta) = \beta.
}
Therefore, this determines $\beta$ and in turn, $\alpha$.
Thus,
\es{}{
\psi &= \eta - {{\cal H} \over 2 k^2} (h' + 6\eta'),\\
\phi &= -{\cal H}\alpha - \alpha' ={{\cal H} \over 2k^2}(h'+6\eta') +  {1\over 2k^2}(h'' + 6\eta'').
}
With these relations we can express the curvature perturbation on uniform density surface as,
\es{}{
\zeta = -\psi - {{\cal H}\over \rho'}\delta\rho = -\eta + {{\cal H}\over 2k^2} (h'+6\eta') - {\cal H \over \rho'}\tilde{\delta\rho}+{\cal H}\alpha = -\eta -{\cal H \over \rho'}\tilde{\delta\rho}.
}
Therefore, requiring $\eta\rightarrow 0$ and $\tilde{\delta\rho}\rightarrow 0$ at early conformal time $\tau\rightarrow 0$ indeed corresponds to vanishing curvature perturbation.

\section{Boltzmann Equations}\label{app:boltzmann_eq}
We describe details of the Boltzmann equation Eq.~\eqref{eq:Boltzmann_hierarchy_decoupling} in this appendix, mostly based on \cite{Brinckmann:2022ajr}. As shown in  Ref.~\cite{Oldengott:2014qra}, for the case of $2\leftrightarrow 2$ scattering process, where we label the momenta of initial (final) particles being $\mathbf q, \mathbf l\,( \mathbf q', \mathbf l')$,  the collision term $C[f]$ that is first order in $\Psi$ is defined as
\bea\label{eq:C[f]}
&&C[f]_{2\leftrightarrow2}(\mk, \mq,\tau)\nonumber\\&&=\frac{g^3}{2 q (2\pi)^5}\int \frac{d^3\ml}{2l}\int \frac{d^3\mq'}{2q'}\int \frac{d^3\ml'}{2l'}\,\delta^4(q+l-q'-l') \langle |\mathcal{M}|^2 \rangle\nonumber\\
&& \qquad \times \left[\bar f^{\rm eq}(q') \bar f^{\rm eq}(l')[\Psi(\mk,\mq',\tau)+\Psi(\mk,\ml',\tau)]-\bar f^{\rm eq}(q) \bar f^{\rm eq}(l)[\Psi(\mk,\mq,\tau)+\Psi(\mk,\ml,\tau)]\right], \nonumber \\
\eea
where $g$ is the degree of freedom of the particle and $\langle |\mathcal{M}|^2 \rangle$ is the matrix element squared averaged over \textit{initial} and \textit{final spins} of all particles with a factor  $1/(N_i ! N_f!)$ that removes the double counting of identical particles. In addition, we include the factor $N_i$ in the matrix element after integrating the Boltzmann equation over $q$ because $N_i$ particles are affected by the interactions considered here.\footnote{In principle, this factor should not be included in Eq.~(\ref{eq:C[f]}), but we just include this factor in $\langle |\mathcal{M}|^2 \rangle$ as we always calculate integrated Boltzmann equations.} From this matrix element squared, we can calculate the thermal averaged rate
\begin{eqnarray}
\langle \Gamma \rangle &\equiv& \frac{1}{\bar n}\int d\Pi_1 d\Pi_2 d\Pi_3 d\Pi_4 \bar{f}(E_1) \bar{f}(E_2) (1 \pm \bar{f}(E_3))(1 \pm \bar{f}(E_4)) \langle |\mathcal{M}|^2 \rangle (2\pi)^4\delta^{(4)}(p_1+p_2-p_3-p_4)\nonumber\\
&\approx& \frac{g^2}{\bar n}\int \frac{d^3 \mathbf p_1}{(2\pi)^3} \frac{d^3 \mathbf p_2 }{(2\pi)^3} \bar{f}(E_1) \bar{f}(E_2)  \sigma_{2 \rightarrow 2}v_{\rm rel} \label{eq:thermalavgrateapprox}
\end{eqnarray}
where $p_i=(E_i,\mathbf{p}_i)$ is the 4-momentum of each particle and $d\Pi_i\equiv\frac{g d^3 \mathbf{p}_i}{(2\pi)^3 2E_i}$ with spin degeneracy $g $, $\bar n\equiv g \int d^3 \mathbf{p}/(2\pi)^3 \bar f(E)$ is the equilibrium number density and we use $n= g N T^3/\pi^2$ for massless particles. The quantity $\sigma_{2 \rightarrow 2}$ is the cross section for $2\to2$ self-interaction and $v_{\rm rel}=s/(2E_1E_2)$ is the relative velocity of initial particles.

To get the analytic expressions for the collision terms Eq.~(\ref{eq:C[f]}) after integration, we need several approximations. First, we assume the background distributions in the comoving frame of all kinds of DR in our analysis are the Maxwell-Boltzmann distributions: $\bar f(q,\tau)=N e^{-q/T_{D,0}}$, where $T_{D,0}$ is the temperature of DR today and $N$ is chosen to match the energy density between the Maxwell-Boltzmann and Bose-Einstein/Fermi-Dirac distributions. Second, we neglect Pauli blocking or Bose enhancement of final state particles.

\subsection{Decoupling }
As an example of decoupling DR, we consider Majorana fermions, which have self-interactions mediated by a heavy mediator (see section~\ref{sec:intDR}). Then the matrix element squared for the self-interaction is $\left< |\mathcal{M}|^2 \right> = \frac{1}{16} G_\phi^2 (s^2+t^2+u^2)$, where $G_\phi$ is the Fermi constant for the interaction and $s,t,u$ are Mandelstam variables. Then, Boltzmann hierarchy for $\Psi$ is given by \cite{Oldengott:2014qra,Oldengott:2017fhy}:
\bea
\dot{\Psi}_0(q) &=& -k \Psi_1(q) + \frac{1}{6}\frac{\partial \ln \bar{f}}{\partial \ln q} \dot{h}-\frac{10}{3} \frac{N T_{D,0}^4 G_\eff^2}{a^4 (2\pi)^3}q \Psi_0(q)\nonumber\\
&& \qquad + \frac{N G_\eff^2}{2 a^4 (2\pi)^3} \int \dd q' \left[ K_0^m(q,q')-\frac{10}{9} q^2 q'^2 e^{-q/T_{D,0}} \right ] \frac{q' \bar{f}(q')}{q \bar{f}(q)} \Psi_0 (q')\label{eq:Boltzmann_hierarchy_decoupling_Psi0}\\
\dot{\Psi}_1(q) &=& -\frac{2}{3}k \Psi_2(q) + \frac{1}{3} k \Psi_0(q)-\frac{10}{3} \frac{N T_{D,0}^4 G_\eff^2}{a^4 (2\pi)^3}q \Psi_1(q)\nonumber\\
&& \qquad + \frac{N G_\eff^2}{2 a^4 (2\pi)^3} \int \dd q' \left[ K_1^m(q,q')-\frac{5}{9} q^2 q'^2 e^{-q/T_{D,0}} \right ] \frac{q' \bar{f}(q')}{q \bar{f}(q)} \Psi_1 (q')\\
\dot{\Psi}_2(q) &=& -\frac{3}{5}k \Psi_3(q) + \frac{2}{5} k \Psi_1(q) - \frac{\partial \ln \bar{f}}{\partial \ln q} \left( \frac{2}{5} \dot{\eta} +\frac{1}{15}\dot{h} \right)-\frac{10}{3} \frac{N T_{D,0}^4 G_\eff^2}{a^4 (2\pi)^3}q \Psi_2(q)\nonumber\\
&& \qquad + \frac{N G_\eff^2}{2 a^4 (2\pi)^3} \int \dd q' \left[ K_2^m(q,q')-\frac{1}{9} q^2 q'^2 e^{-q/T_{D,0}} \right ] \frac{q' \bar{f}(q')}{q \bar{f}(q)} \Psi_2 (q')\\
\dot{\Psi}_{\ell>2}(q) &=& -\frac{k}{2\ell+1}[\ell \Psi_{\ell-1}(q) -(\ell+1)\Psi_{\ell+1}(q)] - \frac{10}{3} \frac{N T_{D,0}^4 G_\eff^2}{a^4 (2\pi)^3}q \Psi_\ell(q)\nonumber\\
&& \qquad + \frac{N G_\eff^2}{2 a^4 (2\pi)^3} \int \dd q' K_\ell^m(q,q') \frac{q' \bar{f}(q')}{q \bar{f}(q)} \Psi_\ell (q') .\label{eq:Boltzmann_hierarchy_decoupling_Psi3}
\eea
Here, $T_{D,0}$ is the temperature of the DR today. The quantity $\bar f$ denotes the averaged phase space distribution and $\Psi_\ell$ is the Legendre decomposition of $\Psi$,
\beq
\Psi(\mk,\mq,\tau) = \sum_{\ell=0}^{\infty} (-i)^\ell (2\ell+1) \Psi_\ell(k,q,\tau) P_\ell(\cos \epsilon) \, ,
\eeq
where $\cos \epsilon = \mk \cdot \mq/(kq)$ and $P_\ell(\cos \epsilon)$ is a Legendre polynomial of order $\ell$. The function $K_\ell^m(q,q')$ is defined as 
\beq
K_\ell^m(q,q') = \int_{-1}^1 \dd \cos \theta K^m(q,q',\cos \theta) P_\ell (\cos \theta),
\eeq
where
\bea
K^m(q,q',\cos \theta) &=& \frac{T_{D,0}^4}{16 P^5} e^{-(Q_- + P)/2}(Q_-^2-P^2)^2 \left[P^2(3P^2-2P-4)+Q_+^2(P^2+6P+12) \right],\nonumber\\
\eea
with $P=|\mq-\mq'|/T_{D,0}$ and $Q_{\pm}=(q\pm q')/T_{D,0}$. 
We further use the relaxation time approximation. This can be achieved by making the following approximation~\cite{Oldengott:2017fhy}:
\bea\label{eq:separable_ansatz}
\Psi_\ell(k,q,\tau)\approx -\frac{1}{4}\frac{d \ln \bar f}{d\ln q} F_\ell(k, \tau).
\eea
Note that $F_\ell$ is defined in Eq.~\eqref{eq:Fell}. We then integrate the Boltzmann hierarchy for $\Psi_\ell$ in Eqs.~(\ref{eq:Boltzmann_hierarchy_decoupling_Psi0}-\ref{eq:Boltzmann_hierarchy_decoupling_Psi3}) to get the Boltzmann hierarchy for $F_\ell$ shown in Eq.~(\ref{eq:Boltzmann_hierarchy_decoupling}). 
With the definition of $\Gamma_\eff$ in Eq.~(\ref{eq:ave_rate_decoupling}), we can find the coefficients $\alpha_\ell$ by matching two equations. We get $\alpha_0 = \alpha_1=0$, which is consistent with energy and momentum conservation. For higher order $\alpha_\ell$, we get $\alpha_2=1.39, \alpha_3=1.48, \alpha_4=1.57$, and $\alpha_5=1.62$.

\subsection{Recoupling}
For the recoupling case, we consider scalar DR with a $\phi^4$ interaction (see section~\ref{sec:intDR}). In this case, the matrix element squared for the self-interaction is $\left< |\mathcal{M}|^2 \right> =\frac{1}{2} \lambda_\phi^2$, and the Boltzmann equations for $\Psi$ read
\bea
\dot{\Psi}_0(q) &=& -k \Psi_1(q) + \frac{1}{6}\frac{\partial \ln \bar{f}}{\partial \ln q} \dot{h} - \frac{N \lambda_\varphi^2 T_{D,0}^2}{128 \pi^3 q} \Psi_0(q)\nonumber\\
&& \qquad + \frac{N \lambda_\varphi^2}{128 \pi^3} \int \dd q' \left[ K_0^0(q,q') - e^{-q{/T_{D,0}}} \right ] \frac{q' \bar{f}(q')}{q \bar{f}(q)} \Psi_0 (q')\label{eq:Boltzmann_hierarchy_recoupling_Psi0}\\
\dot{\Psi}_1(q) &=& -\frac{2}{3}k \Psi_2(q) + \frac{1}{3} k \Psi_0(q) - \frac{N \lambda_\varphi^2 T_{D,0}^2}{128 \pi^3 q} \Psi_1(q)\nonumber\\
&& \qquad + \frac{N \lambda_\varphi^2}{128 \pi^3} \int \dd q' K_1^0(q,q') \frac{q' \bar{f}(q')}{q \bar{f}(q)} \Psi_1 (q')\\
\dot{\Psi}_2(q) &=& -\frac{3}{5}k \Psi_3(q) + \frac{2}{5} k \Psi_1(q) - \frac{\partial \ln \bar{f}}{\partial \ln q} \left( \frac{2}{5} \dot{\eta} + \frac{1}{15}\dot{h} \right) - \frac{N \lambda_\varphi^2 T_{D,0}^2}{128 \pi^3 q} \Psi_2(q)\nonumber\\
&& \qquad + \frac{N \lambda_\varphi^2}{128 \pi^3} \int \dd q' K_2^0(q,q') \frac{q' \bar{f}(q')}{q \bar{f}(q)} \Psi_2 (q')\\
\dot{\Psi}_{\ell>2}(q) &=& -\frac{k}{2\ell+1}[\ell \Psi_{\ell-1}(q) -(\ell+1)\Psi_{\ell+1}(q)]  - \frac{N \lambda_\varphi^2 T_{D,0}^2}{128 \pi^3 q} \Psi_\ell(q)\nonumber\\
&& \qquad + \frac{N \lambda_\varphi^2}{128 \pi^3} \int \dd q' K_\ell^0(q,q') \frac{q' \bar{f}(q')}{q \bar{f}(q)} \Psi_\ell (q')\label{eq:Boltzmann_hierarchy_recoupling_Psil}
\eea
where
\bea
K^0(q,q',\cos \theta) &=& \frac{e^{-(Q_- + P)/2}}{P}.
\eea
To get the Boltzmann hierarchy for $F_\ell$ in Eq.~(\ref{eq:Boltzmann_hierarchy_decoupling}), we integrate Eqs.~(\ref{eq:Boltzmann_hierarchy_recoupling_Psi0}-\ref{eq:Boltzmann_hierarchy_recoupling_Psil}) with the approximation in Eq.~(\ref{eq:separable_ansatz}).  Based on the definition of $\Gamma_\eff$ in Eq.~(\ref{eq:ave_rate_recoupling}), we can get the $\alpha_\ell$ coefficients for the recoupling case: $\alpha_2=0.188, \alpha_3=0.294, \alpha_4=0.356$, and $\alpha_5=0.395$. Again, $\alpha_{0,1}=0$ due to energy and momentum conservation.

\bibliographystyle{utphys-modified}
\bibliography{references}
\end{document}